\documentclass[twocolumn]{aastex62}
\usepackage{amsmath,amstext}
\usepackage[T1]{fontenc}
\usepackage{apjfonts}
\usepackage[figure,figure*]{hypcap}
\usepackage{chngcntr}

\newcommand{\methodpaper}{Paper I}

\newcommand{\fiducialstyle}{black}

\newcommand{\radstyle}{dark blue} 
\newcommand{\ionstyle}{light blue}
\newcommand{\pelwstyle}{blue}

\begin{document}

\title{The Role of Stellar Feedback in the Chemical Evolution of a Low Mass Dwarf Galaxy}

\correspondingauthor{Andrew Emerick}
\email{aemerick@carnegiescience.edu}

\author[0000-0003-2807-328X]{Andrew Emerick}
\altaffiliation{Pasadena Fellow in Theoretical Astrophysics}
\affiliation{Carnegie Observatories, Pasadena, CA, 91101, USA}
\affiliation{TAPIR, California Institute of Technology, Pasadena, CA, 91125, USA}
\author[0000-0003-2630-9228]{Greg L. Bryan}
\affiliation{Department of Astronomy, Columbia University, New York, NY, 10027, USA}
\affiliation{Center for Computational Astrophysics, Flatiron Institute, 162 5th Ave, New York, NY, 10010, USA}
\author[0000-0003-0064-4060]{Mordecai-Mark Mac Low}
\affiliation{Department of Astrophysics, American Museum of Natural History, New York, NY, 10024, USA}
\affiliation{Department of Astronomy, Columbia University, New York, NY, 10027, USA}
\affiliation{Center for Computational Astrophysics, Flatiron Institute, 162 5th Ave, New York, NY, 10010, USA}


\begin{abstract}
We investigate how each aspect of a multi-channel stellar feedback model drives the chemodynamical evolution of a low-mass, isolated dwarf galaxy using a suite of high-resolution simulations. Our model follows individual star particles sampled randomly from an adopted initial mass function, considering independently feedback from: supernovae; stellar radiation causing photoelectric heating of dust grains, ionization and associated heating, Lyman-Werner (LW) dissociation of H$_2$, and radiation pressure; and winds from massive main sequence (neglecting their energy input) and asymptotic giant branch (AGB) stars. Radiative transfer is done by ray tracing. We consider the effects each of these processes have on regulating the star formation rate, global properties, multi-phase interstellar medium (ISM), and driving of galactic winds. We follow individual metal species from distinct nucleosynthetic enrichment channels (AGB winds, massive star stellar winds, core collapse and Type Ia supernovae) and pay particular attention to how these feedback processes regulate metal mixing in the ISM, the metal content of outflows, and the stellar abundance patterns in our galaxy. We find that---for a low-metallicity, low-mass dwarf galaxy ---stellar radiation, particularly ionizing radiation and LW radiation, are important sources of stellar feedback whose effects dominate over photoelectric heating and \ion{H}{1} radiation pressure. However, feedback is coupled non-linearly, and the inclusion or exclusion of each process produces non-negligible effects. We find strong variations with: the star formation history; the ejection fractions of metals, mass, and energy; and the distribution of elements from different nucleosynthetic sources in both the gas and stars.
\end{abstract}

\section{Introduction}

Stellar feedback clearly drives the complex, multi-physics process of galaxy evolution. Feedback regulates the formation of stars by destroying cold, star-forming gas, providing thermal pressure support and energy to prevent gas from cooling and collapsing into star-forming regions, and by driving metal-rich outflows \citep[see][for a recent review]{Zhang2018}, depleting galaxies of the gas required to form more stars. In addition, stellar feedback drives turbulence in the interstellar medium (ISM) of galaxies, which in turn helps to mix new metals from stellar winds and SNe (SNe) in the ISM, and regulates the phases of galaxies' multi-phase ISM. The successes of modern theoretical models of galaxy evolution \citep[see][]{Vogelsberger2020} rely in large part on advances in modelling the complex process of stellar feedback (see \citealt{SomervilleDave2015} and \citealt{NaabOstriker2017} for recent reviews). It has become evident that SNe feedback alone -- a significant source of thermal and kinetic energy and momentum feedback in galaxies -- does not provide a complete model for the evolution of galaxies. This has propelled the search for models of multi-channel stellar feedback which include some combination of stellar winds, stellar radiation, and cosmic rays, in addition to SNe 
However, the complexities in following each of these processes has precluded the development of a fully self-consistent model for multi-channel stellar feedback capable of faithfully reproducing all galaxy properties. 

In pursuing this goal, high-resolution simulations of galaxy evolution capable of resolving the effects of individual feedback processes are required. The small physical sizes, low star formation rates, and relative simplicity of low-mass dwarf galaxies offers a computationally feasible regime to explore different mechanisms of stellar feedback. Simulations of dwarf galaxies have been used extensively in recent literature for this exact purpose. Such models have been used to explore: the efficacy of various mechanisms for depositing the energy and momentum associated with SN feedback \citep{Smith2018b,Hu2019}; the importance of discrete, stochastic sampling of the initial mass function (IMF) in modelling individual SN explosions \citep[e.g.][]{Revaz2016,Su2018,Applebaum2020}; the role of various forms of feedback in driving metal-rich galactic winds and dwarf galaxy self-quenching \citep[e.g.][]{MacLowFerrara1999,RecchiHensler2013,Caproni2017,Robles-Valdez2017}, with recent focus particularly on the lowest mass dwarf galaxies \citep[e.g.][]{Bland-Hawthorn2015,Cashmore2017,Romano2019}; the role of photoelectric (PE) heating in concert with other mechanisms for stellar feedback in regulating star formation and the warm, neutral ISM in dwarf galaxies \citep{Forbes2016,Hu2016,Hu2017}; the importance of ionizing radiation and radiation pressure in driving dwarf galaxy evolution \citep[e.g.][]{WiseAbel2012, Emerick2018a, Agertz2020}; the role in dwarf galaxy evolution of feedback from high-mass X-ray binaries \citep{Artale2015,Garratt-Smithson2019}, momentum injection from resonant Ly$\alpha$ scattering \citep{Kimm2018}, and from cosmic rays \citep[e.g.][more]{Chen2016}; and explorations of how stellar feedback drives turbulent metal-mixing in dwarf galaxies \citep[e.g.][]{Ritter2015,Corlies2018}.

The general conclusions from these studies are that stellar feedback in low-mass dwarf galaxies is effective, driving significant, metal-rich galactic winds. While SNe are perhaps the dominant mechanism by which these winds are driven, these works demonstrate that a feedback phase before the first SN occurs in a star formation event ($\sim$ 40~Myr) is a critical component to a complete model for stellar feedback. Stellar radiation including ionization, ionization heating, radiation pressure, PE heating, and Lyman-Werner (LW) dissociation of H$_2$ and, to a lesser degree, stellar winds, are important sources of this pre-SN feedback. Not only do these processes help regulate star formation, ISM properties, and drive outflows in their own right, but these works demonstrate that they couple non-linearly to SNe feedback, decreasing the typical densities and modifying ISM structures within which SNe explode, increasing their effectiveness. 

In spite of this progress in our understanding of the role of stellar feedback in galaxy evolution, there remain gaps in our knowledge. In particular, previous works comparing feedback effects focus predominantly on the metal properties of galaxies in terms of total metallicities, and not individual elemental abundances. With a better understanding of how stellar feedback gives rise to stellar abundance patterns, comparison between simulations and observations in multi-dimensional chemical abundance space can be better used as an additional discriminator of different feedback models.





This work builds upon the simulations presented in \citet[][hereafter \methodpaper]{Emerick2019} by examining in greater detail the effects of each component of our multi-channel model for stellar feedback. Our fiducial model follows star formation using individual star particles, capturing the feedback from massive star stellar winds, asymptotic giant branch (AGB) winds, \ion{H}{1} and \ion{He}{1} ionizing radiation followed with ray-tracing radiative transfer, \ion{H}{1} radiation pressure, PE heating from far-ultraviolet (FUV) band radiation, H$_2$ dissociation from LW band radiation, and core-collapse and Type Ia SNe.

In \citet{Emerick2018a}, we investigated the impact that stellar ionizing radiation has on regulating star formation and driving outflows in our simulated dwarf galaxy by comparing our fiducial simulation to runs with ionizing radiation feedback absent, or localized to within 25 pc of each massive star. Expanding that analysis significantly, in this work we utilize a suite of seventeen simulations in which we turn on or off each of these processes in order to examine how they drive the evolution of a low-mass ($M_{\rm vir} \sim 10^{9}$), isolated dwarf galaxy. Complimenting previous research in this regime, we pay unique attention to the role these feedback processes have on driving the individual gas-phase and stellar abundance patterns in our simulated galaxy.

\section{Methods} 
\label{sec:methods}
We refer the reader to \methodpaper\ for a detailed description of our numerical methods, initial conditions, and feedback and chemical evolution models. We briefly summarize the key components of these methods most relevant to this work below. 

We follow the evolution of an idealized, isolated, low-mass dwarf galaxy with an initial gas mass of $M_{\rm gas} = 1.80 \times 10^6$~M$_{\odot}$ initialized as an exponential disk with radial and vertical scale heights of 250~pc and 100~pc respectively. This galaxy is embedded in a static, \cite{Burkert1995} dark matter potential with virial mass $M_{\rm vir} = 2.48\times 10^{9}~M_{\odot}$ and and virial radius $R_{\rm vir}~=~27.4$~kpc. This is evolved using the adaptive mesh refinement hydrodynamics code \textsc{Enzo} \citep{Enzo2014}, with a minimum/maximum spatial resolution of 1.8~pc / 921.6~pc in the simulations presented in \methodpaper. Due to computational constraints, we were unable to perform this study at full resolution, and instead adopt 3.6~pc as the maximum resolution. We refer the reader to the resolution studies comparing maximum resolutions of 1.8~pc, 3.6~pc, and 7.2~pc performed in \methodpaper~ and \cite{Emerick2018b}. The results from the 1.8~pc and 3.6~pc resolution simulations are similar for the properties explored here; it is only the 7.2~pc run that shows a significant difference, in large part because feedback is generally underresolved at that resolution.
The grid is refined to maintain a mass resolution of (at least) 50~M$_{\odot}$ per cell, and to ensure that the Jeans length is resolved by at least eight cells. In addition, a three-zone radius region around any star particle that has active feedback (stellar winds or SNe) is refined to the maximum grid resolution.

We use the chemistry and cooling package \textsc{Grackle} \citep{GrackleMethod} to solve a nine species non-equilibrium chemistry model that includes gas-phase and dust H$_2$ formation, a uniform UV background, and localized self-shielding. This galaxy has an initial total metal mass fraction of $5.4 \times 10^{-4}$ \citep[or $0.03 Z_{\odot}$ taking $Z_{\odot}$ = 0.018 from][]{Asplund2009}. We follow the evolution of 15 individual metal species, C, N, O, Na, Mg, Si, S, Ca, Mn, Fe, Ni, As, Sr, Y, and Ba, whose initial mass fractions are initialized to near-zero (10$^{-20}$). Only the total metallicity affects the physics in our simulation, not the individual metal abundances. 

\subsection{Star Formation and Stellar Feedback}
\label{sec:sf feedback}

Our simulation stochastically forms star particles in dense gas ($n > 50$~cm$^{-3}$ in the 3.6~pc resolution runs presented here) by randomly sampling a \cite{Salpeter1955} IMF and depositing individual star particles from 1~M$_{\odot}$ to 100~M$_{\odot}$. For stars above 8~M$_{\odot}$, we follow their \ion{H}{1} and \ion{He}{1} ionizing radiation using the adaptive ray-tracing radiative transfer method of \cite{WiseAbel2011}, and trace their radiation in the LW and FUV bands using an optically thin approximation. These stars eject mass and energy over their lifetimes through stellar winds, and we include mass and thermal energy injection of both core-collapse and Type Ia SNe. Stars below 8~M$_{\odot}$ have no feedback during their lifetime, except mass and energy deposition of their AGB winds at the end of their lives. For stellar winds and SNe, mass, energy, and metals are injected to the grid by mapping a three-cell radius spherical region ($r=3 \times dx = 10.8~$pc) 
to the grid using a cloud-in-cell interpolation scheme. To reduce the significant computational expense of following a continuous source of hot ($T > 10^6$~K), fast ($v \sim 10^{3}$~km~s$^{-1}$) moving gas, we greatly reduce all stellar wind velocities to 10~km~s$^{-1}$. Given this reduction, we cannot make any strong statements as to the role of stellar wind feedback in the evolution of low mass dwarf galaxies. 

Both \ion{H}{1} and \ion{He}{1} ionizing radiation is followed using the adaptive ray-tracing radiative transfer method of \cite{WiseAbel2011}, coupled to the non-equilibrium chemistry and cooling / heating routines in \textsc{GRACKLE}. Stars in our simulation use the \textsc{OSTAR2002} \citep{Lanz2003} grid of O-type stellar models to compute the \ion{H}{1}, \ion{He}{1}, FUV, and LW band fluxes as a function of stellar surface gravity and surface temperature. These latter two quantities, in addition to stellar radius, are taken as a function of mass and metallicity from the \textsc{Parsec} \citep{Bressan2012,Tang2014} grid of stellar models. For stars with stellar properties off of the \textsc{OSTAR2002} grid, we adopt the associated black body flux given the stellar surface temperature. The resulting black body fluxes were adjusted to produce a continuous curve of flux as a function of stellar mass, separately in each band. Rather than adopting fixed \ion{H}{1} and \ion{He}{1} photon energies for each star, we adopt the average photon energy weighted by the associated black body curve in each band, leading to \ion{H}{1} and \ion{He}{1} ionzing photon energies that span the range 13.6--22.5~eV and 25.0--32.5~eV respectively, depending on stellar surface temperature. We refer the reader to Appendix~B of \methodpaper\ which contains plots of each of these quantities. In our fiducial simulations, we include the effects of radiation pressure on \ion{H}{1} but ignore the absorption of ionizing radiation by dust and re-radiation in the infrared.

We assume FUV and LW band radiation are both optically thin, with local (cell-by-cell) attenuation. LW radiation causes H$_2$ dissociation, while FUV radiation leads to PE heating of dust grains. We follow the PE heating models from \cite{Wolfire2003}, but assuming the dust-to-gas scaling with metallicity $Z$ given by \citet{Remy-Ruyer2014}, which shows a significant decline in the dust content at metallicities $Z < 0.1$~Z$_{\odot}$. Our PE heating rate is
\begin{equation}
    \Gamma_{\rm PE} = (1.3 \times 10^{-24} \rm{erg s^{-1} cm^{-3}}) \epsilon n_{\rm H} G_{\rm eff} D
\end{equation}
where $\epsilon$ is an efficiency factor that in detail depends upon $G_o$, temperature, and the electron number density, but which we adopt to scale weakly with $n_{\rm H}$ (see \methodpaper), $D$ is the dust-to-gas ratio, and
\begin{equation}
    G_{\rm eff} = G_o \exp(-1.33\times 10^{-21} D N_{\rm H})
\end{equation}
is the locally-attenuated FUV flux. $G_o$ is the FUV flux normalized to the solar neighborhood \citep{Habing1968} and $D$ is normalized to the solar value, 6.617 $\times 10^{-3}$. This model is similar to that used in both \cite{Forbes2016} and \cite{Hu2017}, with the exception of the treatment of $\epsilon$ and $D$. However, we note that both of their galaxies were at or above the $Z > 0.1$~Z$_{\odot}$ threshold and therefore have a significantly higher (though still low) dust content. We do not account for H$^-$ photodetachment due to the ISRF, which plays an important role in producing H$_2$ in our low-metallicity, low-dust content galaxy. However, we find that this effect is likely subdominant as long as either ionization or LW radiation are followed (see Appendix E of \methodpaper).

\section{Simulations}
\label{sec:runs}

We present the seventeen different simulations run while varying feedback effects in Table~\ref{table:runs}. Each simulation turns on or off various feedback processes as shown in the table. In each case, SNe and stellar winds are included as a pair in each simulation, though again we note that we do not fully capture the effects of stellar winds in our simulations. All runs contain the same total metal enrichment from SNe, massive star stellar winds, and AGB winds. In runs without SN and stellar winds, the mass and metal ejecta from these channels are instead injected at low velocity (10 km~s$^{-1}$) with a thermal energy equal to the stellar surface temperature. All runs with ionizing radiation include radiation pressure on \ion{H}{1}, except when noted. We test the role of radiation pressure by varying its strength with a constant factor, turning it off in RPx0, and increasing it by factors of 2, 5, and 10 in RPx2, RPx5, and RPx10 respectively. The shortrad simulation is the same as the fiducial simulation, but photons are deleted once they have travelled more than 25~pc from their source. This is an attempt to approximate localized prescriptions for ionizing radiation feedback that only deposit energy and ionize gas in an averaged Str\"omgren sphere around a star particle. We examine the effects of stellar ionizing radiation in our high-resolution 1.8~pc simulations in \cite{Emerick2018a}, comparing our fiducial run with a shortrad simulation and a simulation without ionizing radiation (i.e. SN+PE+LW). 

All simulations are restarted from the same output after the initial collapse phase of the galaxy and formation of the first star particles.\footnote{Due to minor updates to this version of \textsc{Enzo} and the master branch of \textsc{Grackle} since publication of \methodpaper, the 3.6~pc fiducial resolution simulation presented here was re-run in its entirety and is not the same as shown in the Appendix of that work.} Therefore, each run starts with the same 38 initial stars, with total stellar mass of 100 M$_{\odot}$. Three of these stars are above the 8~M$_{\odot}$ threshold for radiation and core-collapse SNe, the most massive of which is 11 M$_{\odot}$.


\begin{deluxetable*}{c|c|c|c|c|c|c}
\label{table:runs}
\tablecaption{A list of the feedback physics included in each of our runs. In every case, metal enrichment from SNe and stellar winds are kept fixed. Runs with SNe and stellar winds turned off simply inject the proper elemental abundances at 10~km~s$^{-1}$ at a temperature equal to the stellar surface temperature. The final column lists the final run time of each simulation.}
\tablehead{
\colhead{Run Name} & \colhead{SN} & \colhead{Stellar Winds} & \colhead{Ionizing Radiation} & \colhead{PE Heating + LW Radiation} & \colhead{Radiation Pressure Factor} & \colhead{End Time (Myr)}
}
\startdata
 Fiducial  & Yes & Yes & Yes & Yes       & 1  & 750 \\
 SN+Ion+PE & Yes & Yes & Yes & PE only   & -  & 552 \\
 SN+Ion+LW & Yes & Yes & Yes & LW only   & -  & 626 \\ 
 SN+Ion    & Yes & Yes & Yes & No        & 1  & 750 \\
 SN+PE+LW  & Yes & Yes & No  & Yes       & 1  & 383 \\
 SN+PE     & Yes & Yes & No  & PE only   & - & 355 \\
 SN+LW     & Yes & Yes & No  & LW only   & - & 487 \\ 
 SN-only   & Yes & Yes & No  & No        & -   & 155 \\
 RPx0      & Yes & Yes & Yes & Yes       & 0   & 750 \\
 RPx2      & Yes & Yes & Yes & Yes       & 2   & 704 \\
 RPx5      & Yes & Yes & Yes & Yes       & 5   & 750 \\
 RPx10     & Yes & Yes & Yes & Yes       & 10  & 717 \\
 shortrad  & Yes & Yes & Yes\tablenotemark{*} & Yes      & 1  & 715 \\
 Ion+PE+LW & No & No & Yes & Yes    & 1 & 750 \\ 
 Ion       & No & No & Yes & No    & 1 & 750 \\ 
 PE+LW     & No & No & No & Yes & - & 750 \\
 NoFeed    & No & No & No & No  & - & 60 
\enddata
 \tablenotetext{*}{The shortrad simulation does include full radiative transfer, but deletes photons once they have travelled 25~pc from their source.}
\end{deluxetable*}

\section{Results}

\subsection{Star Formation Regulation}
\label{sec:sfr}

\begin{figure*}
  \centering
  \includegraphics[width=0.98\linewidth]{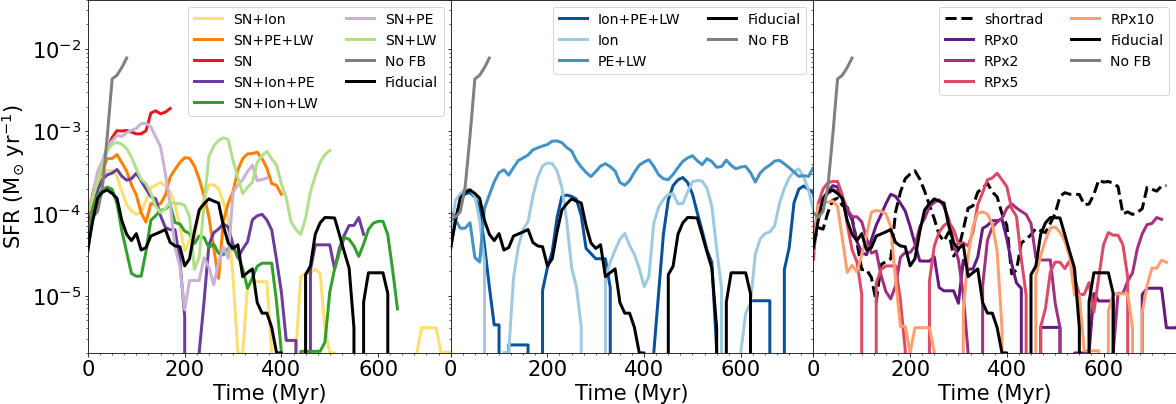}
  \caption{The 50~Myr time-averaged SFR in each of our runs, comparing those with SN feedback varying inclusion of different radiation feedback sources (left), to those without SN (middle), and those with modified radiation pressure and shortrad (right). The fiducial simulation, which includes all physics, is plotted in both panels (\fiducialstyle) for comparison. Feedback from SN, stellar photoionization, and, to a somewhat lesser extent, LW radiation, are all important for regulating the star formation rate.}
  \label{fig:SFR}
\end{figure*}

In Figure~\ref{fig:SFR} we compare the star formation rate (SFR) as a function of time for each of our runs.
Our fiducial simulation shown here (\fiducialstyle) has an initial burst followed by a lower-average, bursty SFR, with extended periods (50~Myr) of no star formation (this is qualitatively similar to the results found in a higher resolution version presented in \methodpaper). The most immediately visible difference between each of the runs is the magnitude of the initial burst of star formation in the first $\sim$100~Myr. Aside from the NoFeed simulation (grey)---which rapidly forms more stars in 50~Myr than the cumulative throughout any of the other simulations---the SN-only run (red) forms the most stars during this time. The peak SFR for each run is ranked quite nicely depending upon the radiation feedback physics included, with PE heating producing the least change from the SN only run, up to ionizing radiation producing the largest change. PE seems to have very little effect on the initial evolution, with LW radiation and ionizing radiation generating the most significant changes. In fact, SN+Ion+LW and the three runs with no SNe that do include either LW or ionizing radiation all have very similar initial bursts of star formation. The only exception to this trend is the shortrad run and Ion+PW+LW, which both have slightly smaller initial bursts. These similarities and the irrelevance of SN feedback on this early phase are clear indications that pre-SN feedback through radiation is important for star formation regulation. 

After the first 100~Myr, the differences between runs due to the inclusion of SN feedback becomes clear. 
Each of the SN-included runs (left panel) show various degrees of bursty star formation for the remainder of the simulation time, with lower average star formation rate than the ionization runs. The run with only PE heating and LW radiation, PE+LW (\pelwstyle), shows the most steady SFR, indicating that this feedback channel alone is capable of producing a self-regulating SFR in this galaxy, but not the bursty star formation that is seen with the inclusion of any other feedback channel. Ionization alone (\ionstyle) and the combined radiation run (\radstyle) both show bursty star formation like the SN runs, but still have a higher average SFR and more consistent SF through the end of the simulation. The SN runs have progressively lower SFR and longer quiescent periods as each simulation proceeds.

The various radiation pressure runs (right panel) do not show striking differences in the evolution of their SFR beyond the types of run-to-run variations expected due to the non-determinstic nature of these simulations. However, it does appear that RPx5 has a slightly elevated SFR for brief period (around 400 Myr) as compared to the other RP simulations. The differences and similarities between these runs in particular will be made more clear in Section~\ref{sec:galaxy properties}.


\subsection{Galaxy Morphology}
\label{sec:morphology}

\begin{figure*}
  \centering
  \includegraphics[width=0.95\linewidth]{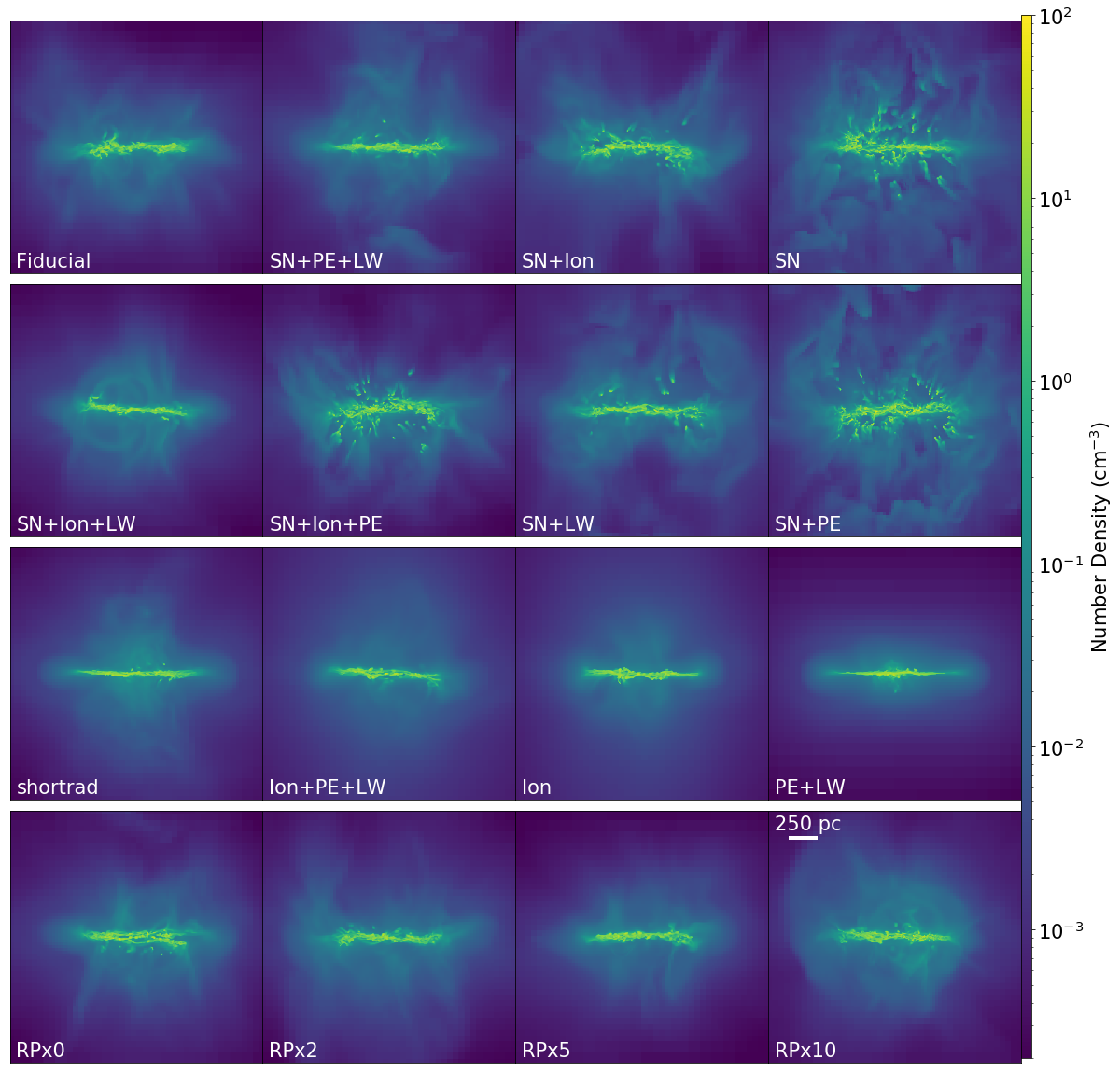}
  \caption{Edge-on projections of gas number density for 
  all of our runs with feedback at time $t=125$~Myr. Feedback from SN tends to drive strong winds from the disk, producing dense clumps above and below the plane; photoionization tends to enhance this effect, but LW radiation tends to diminish the presence of these clumps.}
  \label{fig:panel_plot_1}
\end{figure*}

\begin{figure*}
  \centering
  \includegraphics[width=0.95\linewidth]{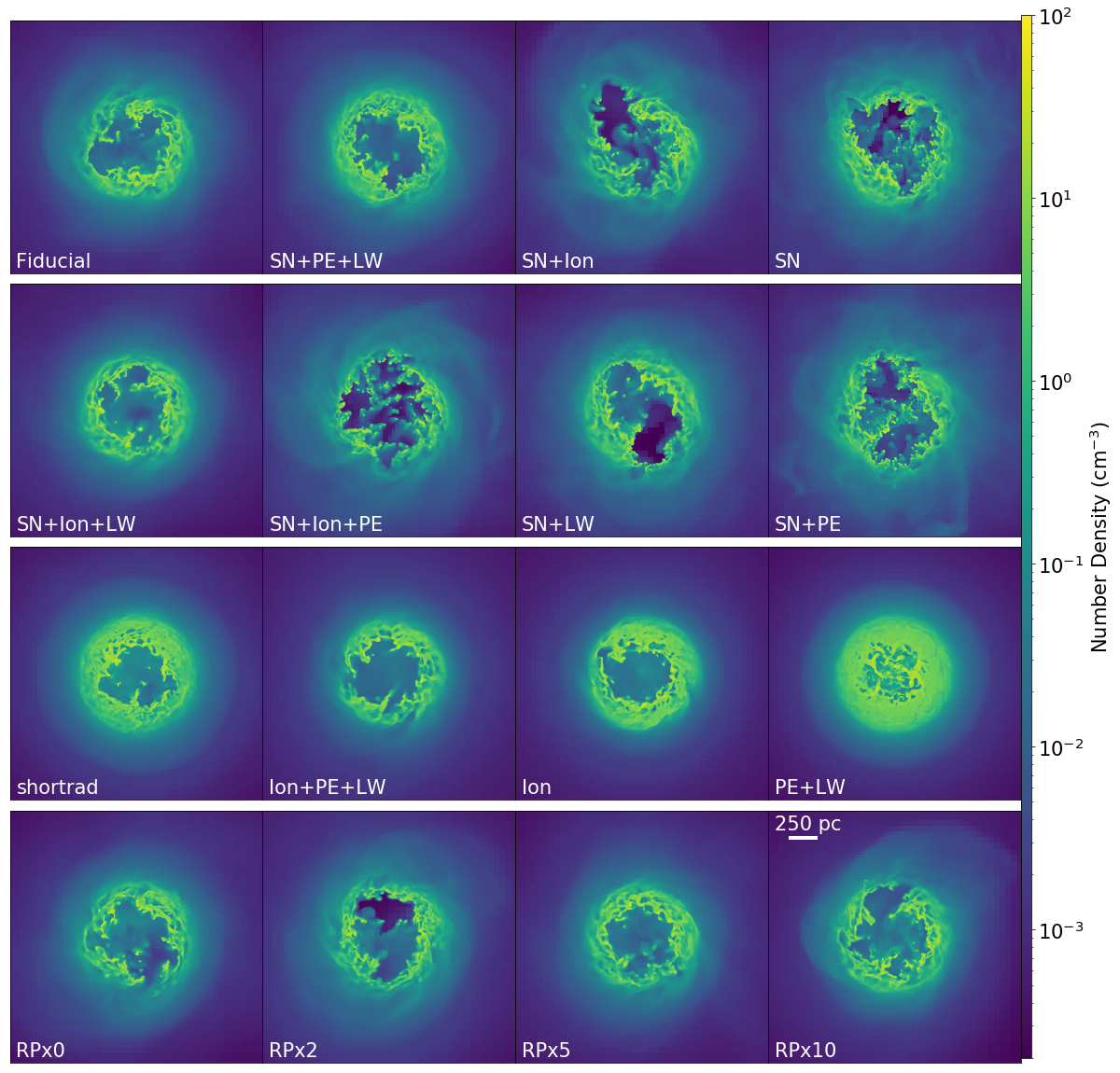}
  \caption{The same panels as Figure~\ref{fig:panel_plot_1}, but face-on.}
  \label{fig:panel_plot_2}
\end{figure*}

We show a comparison of gas morphology of our low-mass dwarf galaxy in edge-on and face-on density-weighted projections of gas number density in Figure~\ref{fig:panel_plot_1} and Figure~\ref{fig:panel_plot_2} respectively at 125~Myr for each simulation except NoFeed (which did not progress to 125~Myr).
In every simulation with SNe, the gas disk is puffier with a larger scale height, and with significant diffuse, outflowing gas above and below the disk. This is not true for the runs with only radiation, which exhibit thin disks with no significant outflowing gas; however ionizing radiation does heat the disk somewhat. The radiation pressure runs (bottom row) do not show significant departures from the fiducial simulation except for RPx10, which appears to have slightly more dense gas and structure beyond the disk.

A subset of these runs show significant extra-planar gas structures, both in the form of diffuse shells of gas and
   poorly resolved, small ($<10$~pc), 
dense clumps of gas above and below the disk (most common in SN, SN+Ion+PE, and SN+PE). Based upon visual analysis of the evolution of these panels, some of the diffuse shells are gas captured in an outflow / re-accretion cycle, while others are gradually pushed out completely. The dense clumps generally originate in the ISM and are entrained in the outflows. These clumps are still bound to the galaxy, and are ablated by the more rapidly moving diffuse gas outflows, giving rise to trailing tails of gas. In some cases these clumps are fully ablated and destroyed, while in others they persist, remain bound to the galaxy, and fall back in on an orbit with larger vertical motions than what is typical of other cold gas in the disk. These clumps do appear at some point in all simulations with either SNe or ionizing radiation, but to a much smaller extent than the obvious cases shown here. Our higher resolution fiducial simulation presented in Paper~I did show some of these features, but again not to the extreme shown here. This suggests that the presence of these clumps may be the resolution dependent, and the result of artificially strong cohesion in under-resolved dense gas \citep{MacLowZahnle1994}. Confirming the physical nature of these clumps will require additional work.

In the face-on panels, each galaxy with SNe or ionizing radiation shows a low-density, carved-out circular region at the center, surrounded by a ring of dense gas. The lack of gas in the center of each galaxy is a particular consequence of the large initial burst of star formation present in every case. However, the stellar feedback from SNe and ionization radiation are (alone and together) effective at heating up and driving out cold ISM in the center of the galaxy as stars form, which leads to a reduction of the central gas content even outside of this burst phase. Only the PE+LW run is incapable of removing gas from the center, maintaining a more uniform and more massive gas disk with localized patches of diffuse gas around newly formed stars. In runs with or without LW and PE (comparing SN+Ion+LW to SN+Ion+PE and SN+LW to SN+PE), LW radiation tends to decrease the presence of dense clumps of gas leading to smaller density contrasts in the inner ISM. As discussed later, this is likely due to the suppression of H$_2$ formation in runs with LW, an important coolant in this low metallicity dwarf galaxy. Many of the clumps seen within the diffuse central region in each simulation are there as projection effects, as they lie mostly outside the mid-plane of the galaxy.  

In Appendix~\ref{appendix:panels}, we show equivalent figures at two other points in time for the available simulations to contrast with this relatively early time in the galaxy's evolution.

\subsection{Global Galaxy Properties}
\label{sec:galaxy properties}

\begin{figure*}
  \centering
  \includegraphics[width=0.95\linewidth]{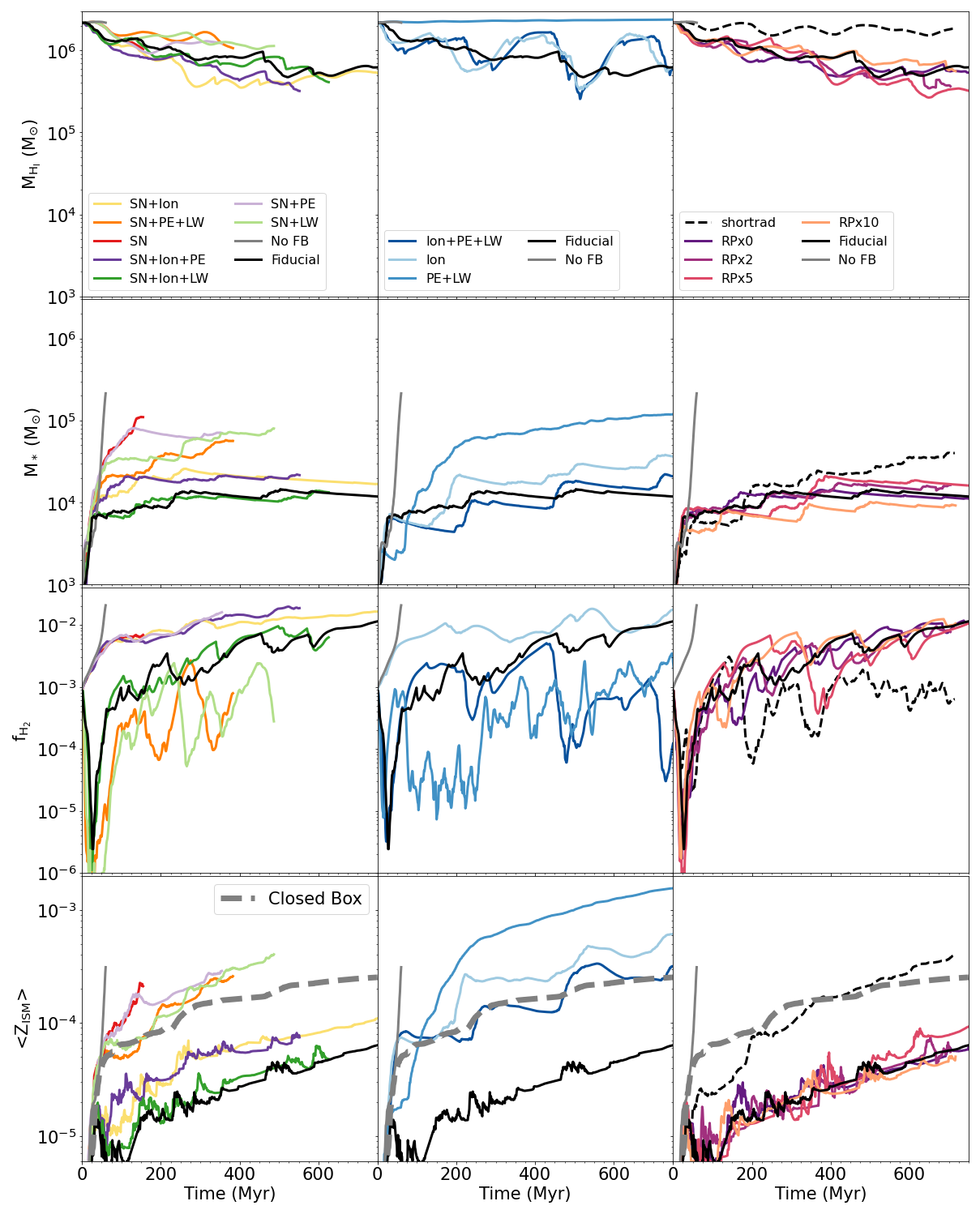}
  \caption{The evolution of four global galaxy properties for each of our simulations. From top to bottom, each row shows: H~{\sc i} mass, stellar mass, H$_2$ fraction, and mean ISM metallicity. $M_{\rm HI}$ and $M_{*}$ have the same vertical axis limits for comparison of the stellar mass fraction between runs. The $\langle Z_{\rm ISM}\rangle $ shown here only includes metals self-consistently produced by stars in this simulation (see text). The HI gas mass is regulated most strongly by the ability to drive outflows, driven by both SN and photoionization. This is true also for the ISM metallicity, which has a non-linear relationship with sellar mass driven by the ability for feedback to drive metals out of the ISM. LW radiation plays the dominant role in regulating the molecular fraction of gas in the galaxy.}
  \label{fig:properties}
\end{figure*}

In Figure~\ref{fig:properties} we compare the time evolution of the total \ion{H}{1}~mass, stellar mass, H$_2$ mass fraction, and average ISM metallicity for each of our runs. The \ion{H}{1}~mass in the fiducial simulation declines with time as stars form, radiation ionizes the ISM, and SN feedback drives significant mass loss. Interestingly the H$_2$ fraction increases significantly during this time, but, as shown in Paper~I, this is mostly due to the preferential retention of the cold, dense gas where the H$_2$ resides, rather than the generation of a significant mass of molecular hydrogen. In the bottom right panel we show the evolution of only those metals self-consistently produced by stars formed in the simulation.
In the fiducial run, the mean ISM metallicity remains well below what could be expected for a closed-box, one-zone model (grey, dashed) with the same star formation history. As discussed in Section~\ref{sec:outflows} this is due to the significant outflows generated by feedback in this galaxy. 

SN+Ion (along with SN+Ion+PE) generally has the lowest \ion{H}{1}~mass of all of the runs. Compared to the fiducial simulation, this is likely due to the increase in stellar mass and associated increases in feedback strength from both SNe and ionization, which both removes gas in outflows and ionizes \ion{H}{1}~in the ISM. The importance of ionizing radiation in driving the \ion{H}{1}~content is seen clearly comparing to the elevated \ion{H}{1}~masses of all runs without ionizing radiation, especially considering that these runs generally have a higher stellar content. The importance of outflows in regulating the gas content can be seen by comparing Ion+PE+LW (\radstyle) to the fiducial, which shows a similar total stellar mass but greater \ion{H}{1}~content, and the ionization only run (\ionstyle) which has a stellar mass higher than both the fiducial and SN+Ion runs but a comparable \ion{H}{1}~mass to the fiducial run. At the extreme, PE+LW radiation alone clearly is incapable of ionizing or removing any \ion{H}{1}~from this galaxy, in spite of its comparatively high stellar mass. Although the shortrad simulation does contain some ionizing radiation, this localized (within 25~pc) ionizing feedback is clearly insufficient to ionize and remove as much gas as many of the other runs. Looking at the runs with or without PE and LW, including PE from SN to SN+PE (or removing it from the fiducial to SN+Ion+LW) provides little difference in any of these properties. However, adding or removing LW radiation does produce a noticeable effect. 

There appear to be three general groupings in the evolution of the molecular fraction between runs. Runs without LW exhibit the highest H$_2$ fractions with the smoothest evolution. The remaining runs with LW result in overall lower and more variable H$_2$ fractions. The runs with the lowest H$_2$ content include LW but are missing another form of feedback, leading to increased stellar masses and an increase in the total LW flux in the galaxy. 


Differences in stellar mass evolution and gas content in these runs lead to significant variations in the mean ISM metal fraction. The fiducial run, RPx0, and SN+Ion+LW show the lowest mean ISM metallicities, a factor of $\sim$4 below that of the closed-box model with the same SFH as the fiducial simulation. The remaining simulations show increasing ISM metallicity, driven by a combination of both increased star formation and decreased outflows that allow greater metal retention. This interplay can be inferred from the fact that the relative differences in $\langle Z_{\rm ISM}\rangle $ between the various runs and the fiducial simulation far exceed the comparable differences in total $M_{*}$. An extreme example of this is comparing PE+LW with SN+PE and SN+LW. All three have similar total stellar masses by 500~Myr, but PE+LW has a factor of several higher metallicity than the other two. We will discuss outflows in more detail in Section~\ref{sec:outflows}.

The radiation pressure runs do vary in each of these quantities, but appear to be scattered around the result from the fiducial simulation, rather than having a consistent trend as a function of radiation pressure factor. Perhaps in line with the assumption that radiation pressure has some effect, RPx10 forms the least stars of this set, but RPx5 forms the most, and RPx0 and RPx2 do not deviate significantly from the fiducial simulation. While how stellar feedback affects star formation in local regions within our galaxy may be affected by radiation pressure -- which could be reflected in some of the variations in this figure -- we do not find any conclusive evidence that radiation pressure as boosted by a factor of up to 10 over the single-scattering limit strongly influences the evolution of our particular low-mass dwarf galaxy. This point is discussed further in Section~\ref{sec:radiation pressure}.

\subsection{Multi-Phase Gas}
\label{sec:phase diagrams}

\begin{figure*}
  \centering
  \includegraphics[width=0.95\linewidth]{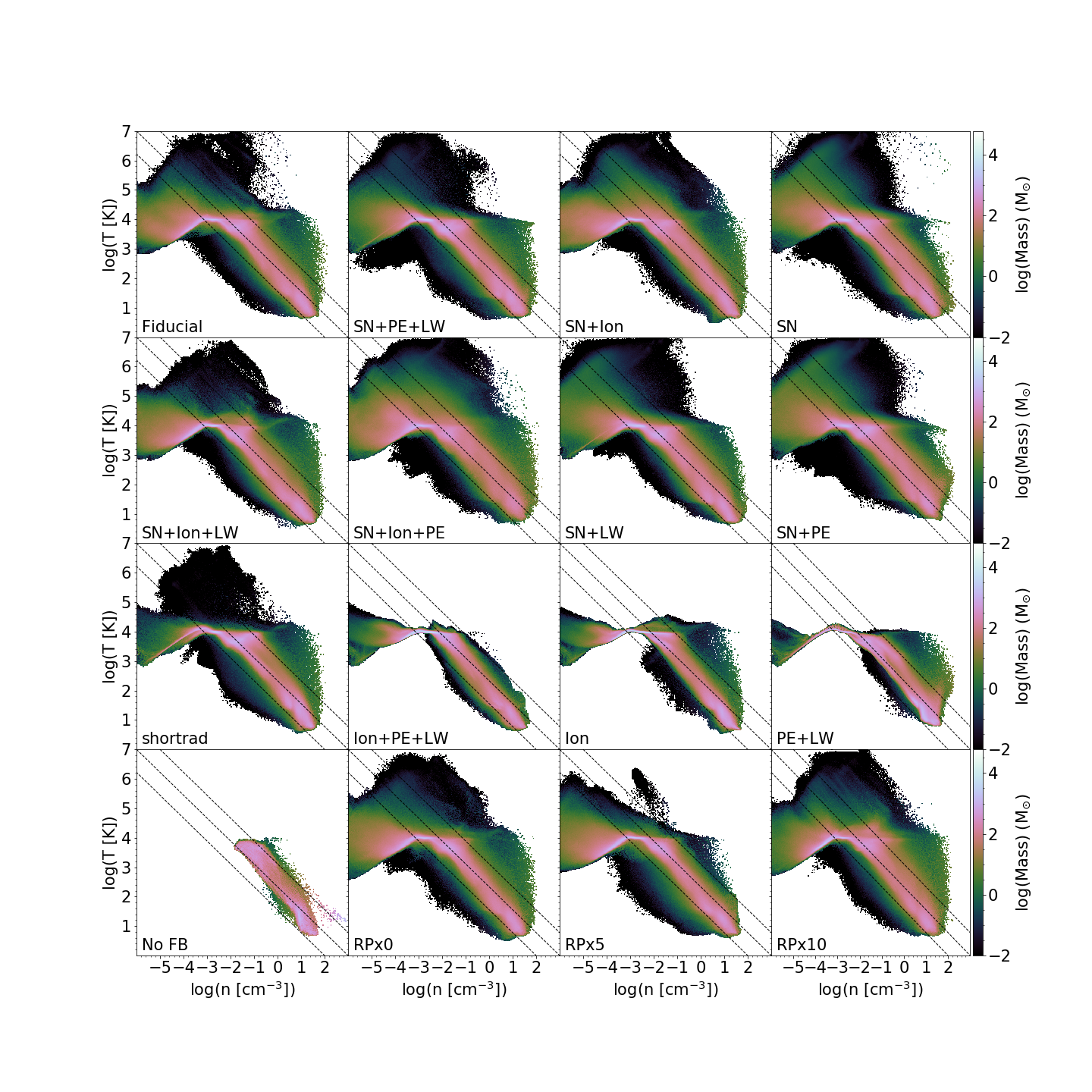}
  \caption{The temperature - density phase diagrams for each of our simulations averaged over the 20~Myr period from 100-120~Myr. For brevity, we choose to exclude RPx2 from this plot. Lines of constant pressure are given as dashed lines. SN heating is required to produce a significant hot phase ($T> 10^5$ K); photoionization feedback results in a warm ionized phase (at $n < 1$ cm$^{-3}$ and $10^4$ K $< T < 10^5$ K).}
  \label{fig:phase_diagram}
\end{figure*}

\begin{figure*}
  \centering
  \includegraphics[width=0.95\linewidth]{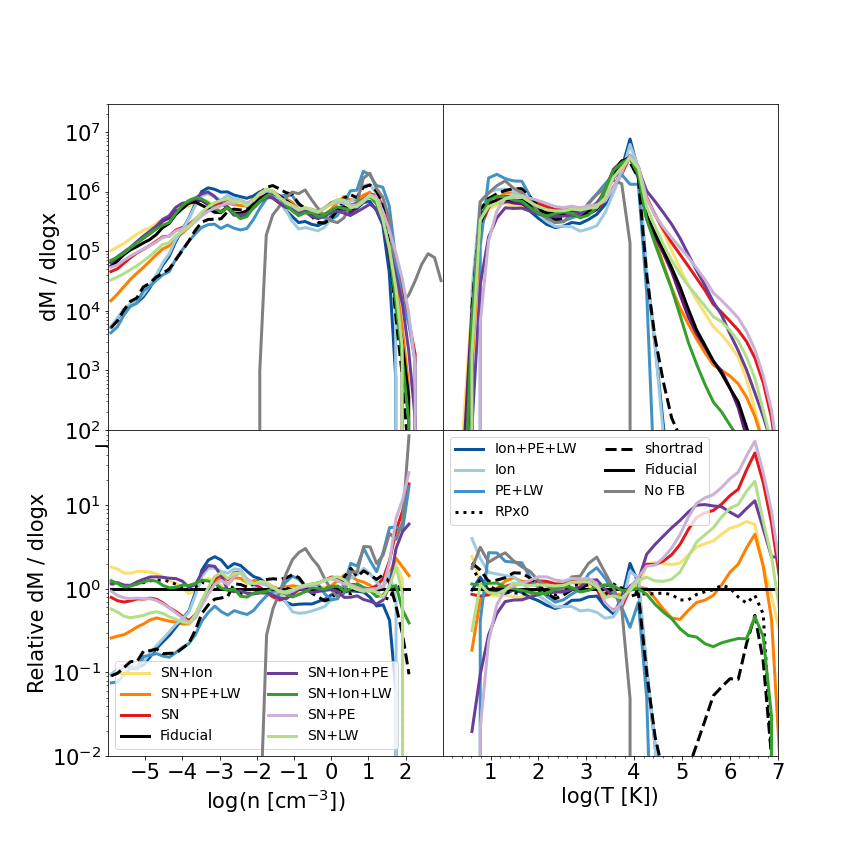}
  \caption{The 1D density and temperature PDFs (top) corresponding to Figure~\ref{fig:phase_diagram}, along with each PDF relative to the fiducial simulation (bottom). For clarity, we do not include RPx2, RPx5, or RPx10 here, which do not show significant variations from the Fiducial run.}
  \label{fig:1D_phase}
\end{figure*}

The properties of the multi-phase ISM for our galaxy simulations are primarily feedback-driven, as shown in the temperature-density phase diagrams (Figure~\ref{fig:phase_diagram}), and the one-dimensional (1D) temperature and density probability density functions (PDFs; Figure~\ref{fig:1D_phase}). Both figures show only gas contained within the disk of each galaxy, and are averaged over a 20~Myr period from 100--120~Myr in each simulation. As indicated by the two-dimensional (2D) phase diagrams, the most obvious differences are in the absence of hot gas at $T > 10^5$~K driven by SNe in the radiation-only runs and the absence of diffuse, warm ($T \sim 10^4$~K) gas at $n < 10^{-2}$~cm$^{-3}$ in the run without any feedback. 

Comparing the runs with and without ionization, it is also clear that this feedback component is necessary to maintain a significant amount of gas in the warm, diffuse phase ($10^4~ {\rm K}~ T < 10^5~ {\rm K}$, $n < 1$~cm$^{-3}$). This is most obvious comparing the runs without SN, but the SN+PE+LW and SN-only runs also have somewhat less gas in this regime. Surprisingly, while shortrad does have ionizing radiation feedback, its limited physical extent also reduced the amount of warm, ionized gas compared to simulations with full ionizing feedback. PE and LW radiation seem to have only slight, subtle effects on the gas phases in these figures. However, comparing SN+LW to SN+PE, we see that PE leads to slightly more cold, dense gas. 

Overall, we see that, generally speaking, including additional feedback sources tends to (overall) broaden the distribution of gas with different densities or temperatures at fixed temperature or density. In detail, we find that SNe are necessary to generate hot gas, but are somewhat less efficient at sustaining warm, ionized gas. The production of that phase requires ionizing radiation---and more than just short-range ionization and heating.

These differences are more distinct in the 1D PDFs shown in Figure~\ref{fig:1D_phase}, giving the full PDFs in the top row and the PDFs normalized to the fiducial simulation in the bottom. Again, runs without SNe lack hot gas, which instead piles into the warm phase right around 10$^4$~K. These same runs also have much more cold, dense gas, as the lack of SNe makes it challenging to destroy cold gas in the ISM. As suggested previously, LW radiation is important for regulating the presence of cold, dense gas, with SN+PE and SN+Ion+PE having elevated cold gas compared to their counterparts with LW instead of PE.  

\subsection{Outflows}
\label{sec:outflows}

Galactic outflows are a natural consequence of stellar feedback, and have been demonstrated to be significant in simulations of low-mass dwarf galaxies \citep[e.g.][]{Ma2016,Muratov2017,Christensen2018,Hafen2019}. While observations of outflows at this galaxy mass are challenged, they are expected to be significant based upon the metal retention fractions of low-mass dwarf galaxies \citep[e.g.][]{McQuinn2015,Zheng2019}, however current observations suggest that simulations may overestimate these outflows \citep[e.g.][]{McQuinn2019}.

\subsubsection{Diagnostics}

The efficiency with which stellar feedback drives outflows is often characterized using the mass loading factor, $\eta_M$, metal mass loading factor, $\eta_Z$, and energy loading factor $\eta_E$. We define each of these quantities following \cite{LiBryan2020}: $\eta_M \equiv \dot{M}_{\rm out} / \dot{M}_{\rm *}$, $\eta_Z \equiv \dot{M}_{Z,\rm{out},\rm{SN}} / \dot{M}_{Z,\rm{SN}}$, and $\eta_E \equiv \dot{E}_{\rm out}$. By definition, $\eta_Z$ is always $<$~1 and represents the fraction of SN-synthesized metals that go into outflows. There is some variety in how exactly $\eta_M$ is computed in simulations, even with this definition. $\dot{M}_{\rm out}$ is the instantaneous mass outflow rate, but $\dot{M}_{\rm *}$ is rarely taken to be the instantaneous star formation rate, often time-averaged on anywhere from 1 to 100~Myr. Particularly for galaxies with bursty star formation, as is the case here, exactly how $\dot{M}_{\rm *}$ is averaged can lead to a large variability in $\eta_M$. To smooth over these fluctuations, we use 100~Myr time-averaged $\dot{M}_{\rm *}$ and compute $\eta_M$, $\eta_Z$, and $\eta_E$ every 5~Myr in each simulation for both the hot ($T \geq 3 \times 10^5$~K) and cool ($T < 3\times10^5$~K) phases. 

\subsubsection{Outflow Strengths}
\label{sec:outflow strengths}
\begin{figure*}
  \centering
  \includegraphics[width=0.825\linewidth]{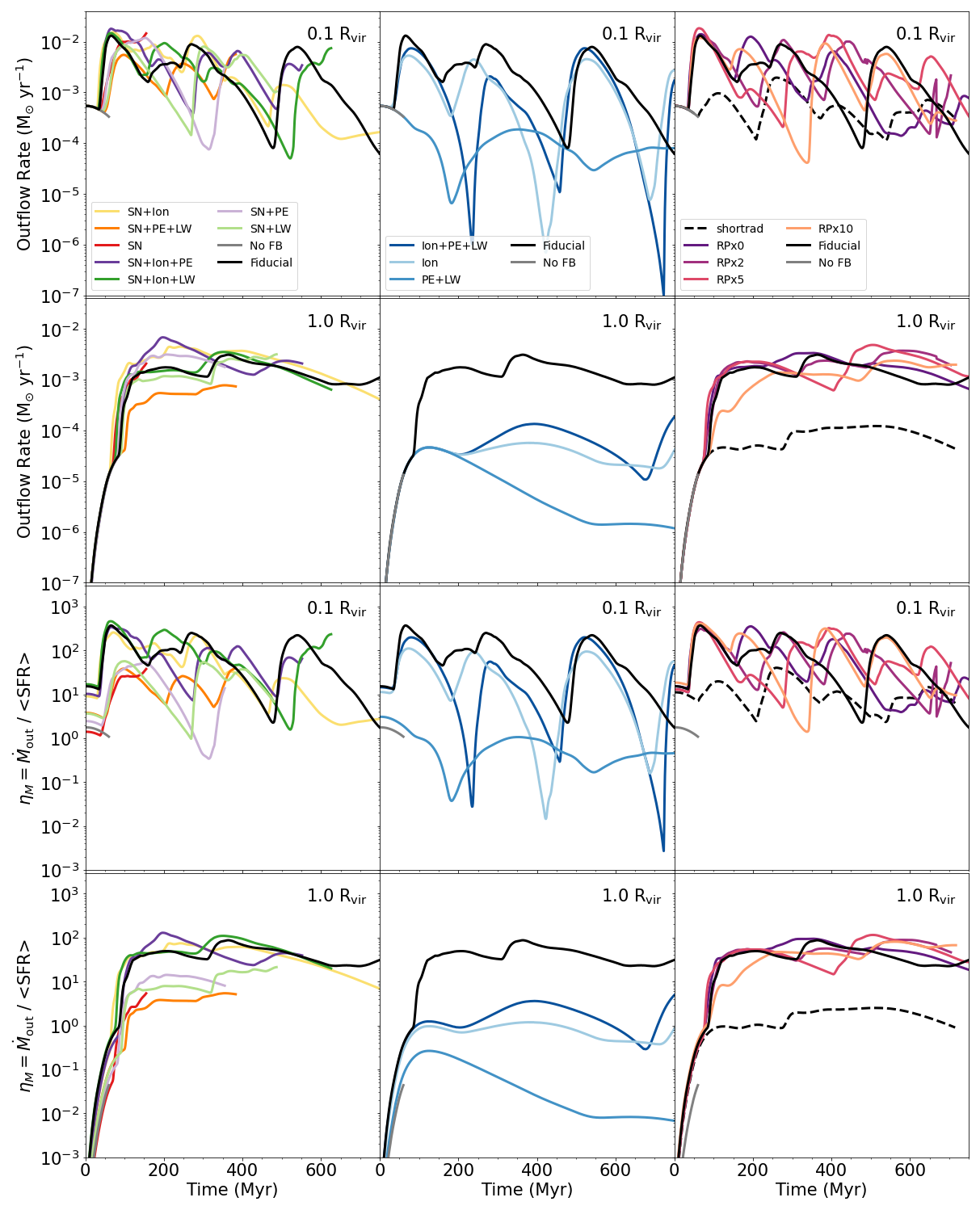}
  \caption{The total mass outflow rate (top two rows) and mass loading factor (bottom two rows) measured through spherical shells at 0.1 and 1.0 R$_{\rm vir}$ for each of our runs except RPx2, RPx5, and RPx10.}
  \label{fig:outflow_evolution}
\end{figure*}

We measure the mass outflow rate through a spherical annulus with thickness 0.05~R$_{\rm vir}$ centered at two radii from the galaxy, 0.1~R$_{\rm vir}$ and R$_{\rm vir}$ (again, R$_{\rm vir}$ = 27.2~kpc). The total outflow rates in the inner halo (Figure~\ref{fig:outflow_evolution}) are within factors of a few of each other during the first $\sim$200~Myr for all of the simulations except shortrad and PE+LW, which are unable to drive significant outflows. Clearly, some combination of full stellar ionizing radiation and/or SNe are necessary to drive outflows in low mass dwarf galaxies. 

After this point, the runs differentiate, but generally fluctuate between similar minimum and maximum values. Runs with SN feedback and ionizing radiation tend to maintain a higher, more consistent outflow rate. The equivalent mass loading factors $\eta_M$ for these runs at 0.1~$R_{\rm vir}$ can be up to 400 and typically $\sim 100$ for all runs including ionizing radiation, and are generally lower (up to 50, but typically $\sim 20$) for those without (and below 1 for PE+LW). This shows that ionizing radiation is important for driving significant outflows, acting on concert with SNe; and in the case of this low mass dwarf galaxy, ionizing radiation alone is capable of driving outflows with $\eta_M$ up to 100 near the disk of the galaxy. It is somewhat surprising that runs without SNe can still reach these large values, but it is the case that 0.1$R_{\rm vir}$ represents a small distance from our galaxy center (2.7~kpc). Only runs with both SNe and ionizing radiation are capable of driving substantial outflows ($\eta_M > 10$) at the virial radius; PE and LW alone have little impact on the outflows. In agreement with the findings in \citet{Emerick2018a}, the implementation of the ionizing radiation feedback is important, as shown by the low $\eta_M$ at $R_{\rm vir}$ in the shortrad simulation. Finally, each of the runs with SN, with the exception of SN+PE+LW, show remarkably similar total mass outflow rates at $R_{\rm vir}$.

The for RP runs again do not show variations that point conclusively to the influence of radiation pressure in either mass outflow or mass loading at either of the two radii examined. However, from RPx0 to RPx10 (and including the fiducial run) there is a slight trend in the full time-average mass loading factor at 0.1 R$_{\rm vir}$ across the runs (from $\eta_M$ = 87 in RPx0 to $\eta_M$ = 106 in RPx10), but this trend is weak (and somewhat inverted) at 1.0 R$_{\rm vir}$. This trend is driven by the radiation pressure on neutral gas external to the disk of the galaxy; an effect which declines at larger radii towards $R_{\rm vir}$.

\subsubsection{Loading Factors}
\begin{figure*}
  \centering
  \includegraphics[width=0.95\linewidth]{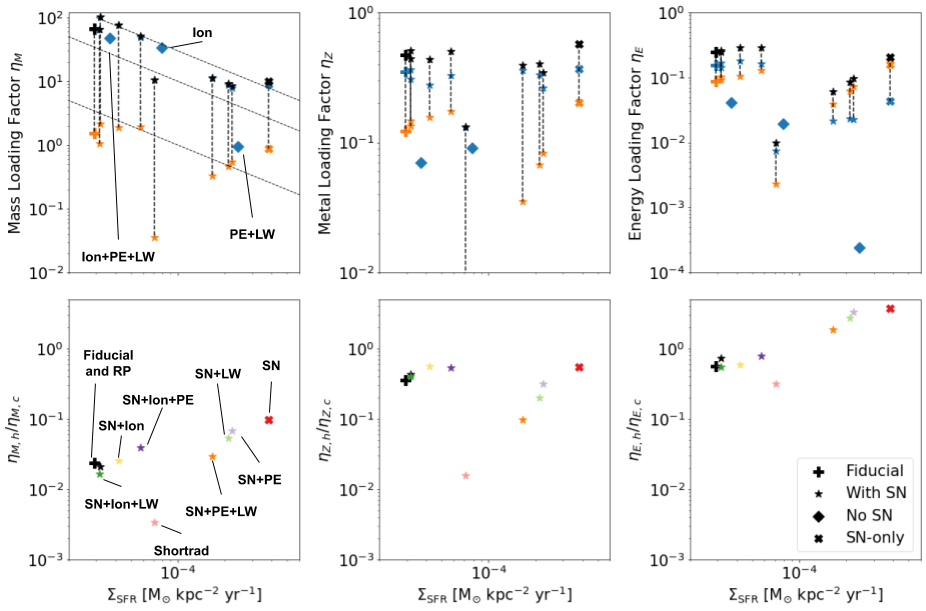}
  \caption{The mass loading factor ($\eta_M$), metal loading factor ($\eta_Z$), and energy loading factor ($\eta_E$) for both hot (orange, $T \geq 3 \times 10^5$~K) and cold (blue, $T < 3 \times 10^5$~K) gas and their ratios for each of our simulations (except NoFeed, RPx2, RPx5, and RPx10) computed at a height of 1~kpc above and below the disk of our galaxy. The totals are shown in the top panel in black, except for the runs without SNe which contain no hot outflows and are left as blue. Each simulation is labelled in the left column, but for clarity our fiducial simulation is shown with a plus, simulations with SNe feedback as stars, those without SNe with diamonds, and SN-only with an X. In order of increasing $\Sigma_{\rm SFR}$ the runs without SNe (visible only in the top panels) are Ion+PE+LW, Ion, and PE+LW. The runs with SNe (visible in all panels) are: Fiducial, SN+Ion+LW, the runs with varying RP, SN+Ion, SN+Ion+PE, Shortrad, SN+PE+LW, SN+LW, SN+PE, and SN.}
  \label{fig:loading_factors}
\end{figure*}

We present the average of each loading factor for each simulation (except NoFeed, RPx2, RPx5, and RPx10) over their entire run-time in Figure~\ref{fig:loading_factors}, as a function of their average star formation rate density. For consistency with the works compared in \citet{LiBryan2020}, we compute these averaged outflow quantities at a height of 1~kpc above and below the disk of our galaxy, as compared to the radial shells in \ref{sec:outflow strengths}.
In general, our simulations show significant amounts of mass and energy driven out via cooler outflows, as the cool gas loading factors equal or far exceed the hot loading factors. This is qualitatively different from the suite of simulations presented in \cite{LiBryan2020} \citep[see also ][]{CGKim2020}
where the hot phase carries at least as much mass and energy out as the cold phase. However, although there is some overlap in $\Sigma_{\rm SFR}$ between our galaxies and a few of the simulations considered in those works, our galaxy has a much lower total mass and gravitational potential. Since the virial temperature of our halo is only $1.4 \times 10^4$~K, it is not surprising that both mass and energy can flow out of our galaxy without reaching temperatures above 10$^5$--10$^6$~K. This is further shown by the fact that the two runs with ionization but no SNe have larger mass loading factors driven only by cold gas. Comparing each run, only the runs with SNe contain any hot outflows. 

In general, there is a decreasing trend of $\eta_M$ with increasing star formation rate. However, this trend generally follows the line of constant $\dot{M}_{\rm out}$ for each simulation, suggesting that each galaxy sustains a similar total mass outflow rate, but it is just the efficiency of stellar feedback in driving those outflows that changes by including or excluding different modes of feedback. As shown above, runs with both ionizing radiation and SNe produce the most efficient outflows, ejecting more gas per solar mass of star formation than runs without either source of feedback. The SN only run has the highest metal loading (slightly), but at a lower mass and slightly lower energy loading than achieved by also including ionizing radiation. The PE+LW simulation generates the least efficient outflow (near unity $\eta_M$, no metal outflow, and very small $\eta_E$).

It is again interesting to note the behavior of the shortrad simulation here. It has one of the lowest mass loading factors of runs with SNe, and metal and energy loading factors more similar to runs without SNe, in spite of the fact that it includes both SNe and ionizing radiation. As was seen in \citet{Emerick2018a}, the ionizing radiation limits the local star formation in this galaxy, but does not punch through the diffuse, ionized channels of gas needed for SNe to effectively drive out mass, metals, and energy. The hot gas is trapped, leading to the very low hot loading factors across all three quantities. In runs without ionizing radiation, this is compensated for with extra SN feedback as the result of additional star formation.




The fraction of SN-produced metals that flow out relative to the total produced $\eta_Z$ is fairly uniform for all runs with SNe ($\eta_Z > 0.5$), but significantly lower for those without SNe ($\eta_Z \lesssim 0.1$), and zero for the PE+LW-only run. Comparing to the differences across simulations in $\eta_M$, this raises two important points: 1) while ionizing radiation alone can drive outflows in the inner-halo of this low-mass dwarf galaxy, SNe are necessary to drive metal-enriched outflows, 2) the metal content of SNe-driven outflows in this low-mass dwarf galaxy is less sensitive to additional feedback processes than the total outflows. Cool outflows carry a majority of the metals out of the ISM of our galaxy in each simulation (as is the case for $\eta_M$).

The energy ejection fraction $\eta_E$ shows similar trends, except that the differences in energy content between hot and cool outflows are much smaller for some of the runs, and for others $\eta_{E,h}$ exceeds $\eta_{E,c}$ by factors of a few. There appears to be a transition $\Sigma_{\rm SFR}$ around 10$^{-4}$~M$_{\odot}$~kpc$^{-2}$~yr$^{-1}$ above which hot outflows dominate over cold in carrying energy. 

\begin{figure*}
  \centering
  \includegraphics[width=0.95\linewidth]{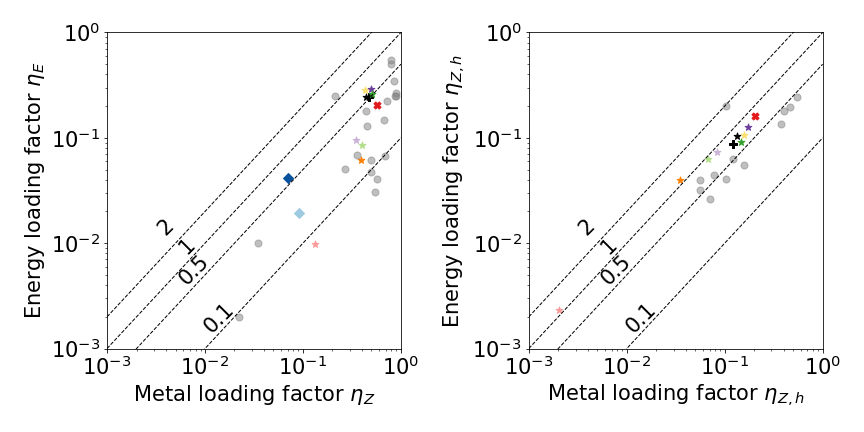}
  \caption{A comparison of the total and hot energy and metal loading factor relationships for our simulations (black points) as compared to the suite of simulations presented in \citet{LiBryan2020} (grey points), which includes data from \citet{Li2017,KimOstriker2018,Fielding2018,Hu2019,Armillotta2019,Martizzi2016} and \citet{Creasey2015}. Lines of constant $\eta_E / \eta_Z$ are shown. We find a relations between energy and metal loading factors that are consistent with, but slightly different from, those found in \citet{LiBryan2020}.}
  \label{fig:loading_relation}
\end{figure*}

Finally, \citet{LiBryan2020} find a strong correlation between the $\eta_{E,h}$ and $\eta_{Z,h}$ across their examined simulations. We plot the relationship between $\eta_{E}$ and $\eta_{Z}$ and $\eta_{E,h}$ and $\eta_{Z,h}$ in Figure~\ref{fig:loading_relation} (black points) as compared to the simulations presented in \citet{LiBryan2020} (grey points). We find that there is no strong correlation between the energy loading factor and metal loading factor (left panel) in our simulations than in \citet{LiBryan2020}; the total energy loading factor is more sensitive to the varying feedback physics than the total metal loading factor. In general, increasing the sources of feedback (particularly SNe and ionization) increases the energy loading factor. Although our simulations without SNe do not contain any hot outflows, the right panel shows that our simulations exhibit a similar linear relationship between $\eta_{E,h}$ and $\eta_{Z,h}$, but at a slightly higher value ($\sim$1) than in \citet{LiBryan2020} ($\sim$0.4).

\subsection{Evolution of Individual Metals}
\label{sec:metals}

Metals released into the ISM in core collapse SNe are ejected from the galaxy through outflows with a higher efficiency than metals from AGB winds, leading to a lower fraction retained in the disk, as explored in \citet{Emerick2018b}. These same metals show smaller gas-phase abundance variations---pointing to more efficient mixing---than elements from AGB winds. We explore how different feedback mechanisms affect these differences here.

\begin{figure*}
  \centering
  \includegraphics[width=0.95\linewidth]{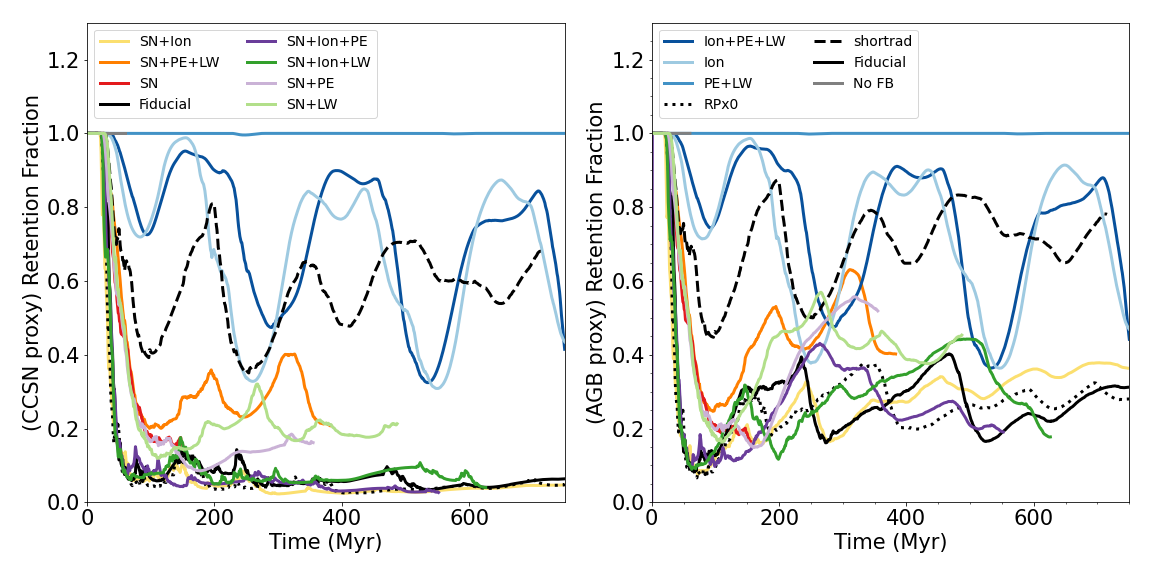}
  \caption{The retention fraction of CCSN elements (left, traced by O) and AGB elements (right, traced by Ba) in the disk of each galaxy as a function of time. For clarity, we do not show RPx2, RPx5, or RPx10 here. Elements generated in CCSN (as opposed to those from AGB) are more efficiently ejected by simulations including the energetic input from those SN, but simulations with other forms of feedback do not show this difference
  \label{fig:retention}}
\end{figure*}

In Figure~\ref{fig:retention} we plot the fraction of metals produced within the disk at each point in time for a core-collapse supernova (CCSN) proxy (O, left)\footnote{We note that this quantity does not vary significantly for other SN-dominated elements, like Mg.} and an AGB enrichment proxy (Ba, right). The fiducial simulation shows a fairly consistent CCSN retention fraction of $\sim$5\% after the first 100~Myr. The AGB retention fraction drops during the first 100~Myr until the first AGB stars begin producing significant Ba; the retention fraction then grows and oscillates with the SFR, but remains between 20\% and 40\%, significantly higher than that for CCSN elements. 

In general, runs with both SN feedback and ionization show low CCSN element retention. Runs without ionization (SN+PE, SN+LW, SN+PE+LW) retain a much larger fraction of their CCSNe elements (20--40\%), but still have slightly higher AGB-element retention fractions (40--60\%). Only runs without SNe show similar retention fractions across both CCSNe and AGB elements. The large fluctuations in the shortrad, Ion, and Ion+PE+LW runs are caused by gas that is removed beyond our formal definition of the disk region, but is not ejected far into the halo and eventually re-accretes onto the galaxy. This is additional confirmation that this difference is due in large part to the difference in energetics between the SN and AGB events, as expected from the analysis in \citet{Emerick2020a}. Retention fractions are similar for different elements if the mode of outflows is dominated by swept-up ISM rather than the direct release of metals from the enrichment source itself (in this case, SNe).


\begin{figure*}
  \centering
  \includegraphics[width=0.95\linewidth]{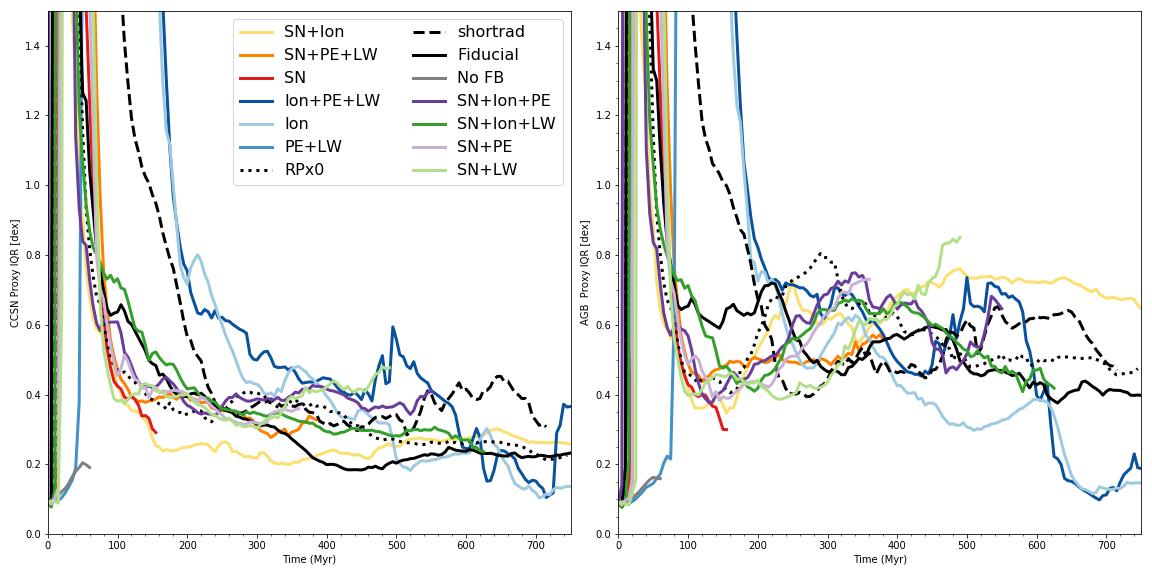}
  \caption{A comparison of the evolution of the IQR of the gas-phase abundances of CCSN elements (left, traced by O) and AGB wind elements (right, traced by Ba) in the cold ISM ($T < 10^2$~K).For clarity, we do not show RPx2, RPx5, or RPx10 here.  Run PE+LW (\pelwstyle) remains well above the vertical bounds for the duration of the simulation, and is omitted for clarity (see text).}
  \label{fig:mixing}
\end{figure*}

Finally, we explore the efficiency of mixing of these elements these in our suite of simulations in Figure~\ref{fig:mixing}. We compare the interquartile range (IQR) abundance spread of the cold gas for O, the CCSN element proxy, and Ba, the AGB element proxy, for each run. In the fiducial simulation, the CCSN elements show a declining spread throughout the evolution, reaching an IQR of $\sim$0.25 dex by the end of the simulation time. As was seen in the retention fraction, the AGB elements follow the same evolution until the point in time when they are first produced by AGB rather than CCSNe, at which point the width rises significantly, to around 0.7~dex, and remains well above the CCSN elements until the end of the simulation, though it does decline to about 0.4~dex. All of the simulations including SN feedback show very similar general trends and final IQR values for O, with the exception of shortrad which takes twice as long to reach the initial plateau at about 0.4~dex, and stays above the fiducial simulation for the remainder of the time. The comparison here and to the radiation-only runs shows the importance of hot-phase mixing to the evolution of these elements. While the runs with SNe hit a roughly consistent IQR after the first $\sim$150~Myr, the simulations without SNe take more than twice as long to reach this point. This happens despite the radiation-only runs having a higher global SFR with more continual and uniformly distributed enrichment that should allow for more rapid homogenization. However, the radiation-only runs do not have SNe first depositing their elements in a volume-filling hot phase, which is necessary for rapid mixing over the whole galaxy \citep{Emerick2018b}. Ionizing radiation does generate enough warm gas to increase the mixing timescales over what is present in PE+LW, which is not shown in either panel for clarity. PE+LW exhibits mixing timescales on order of or exceeding the dynamical time of the galaxy ($\sim$1~Gyr); by 300~Myr, this run has an IQR of 10~dex for both elements, declining to 2~dex by 750~Myr in both cases.

Since SNe and AGB winds have the same injection energy in the radiation-only runs, they exhibit similar abundance spreads. This suggests that, not only is the mean abundance of individual elements an important discriminator between feedback models, but also the scatter in individual abundances can contain information about the stellar feedback that drives metal mixing in the ISM. Future work correlating gas-phase abundance spreads with ISM properties for a range of feedback models would be useful in making the connection from observed gas-phase abundances to stellar feedback parameters.

Although the circumgalactic medium of a galaxy of this size will be challenging to observe directly, the presence of metals ejected beyond the disk of the galaxy may also be an important discriminator between feedback models. We calculated the CGM metal content, but for brevity the associated figure is not shown. We find that the fraction of metals in the CGM for CCSN is initially large after the first burst of star formation ($\sim$80--90\%) for all SN runs, but gradually declines to $\sim$20--30\% by the end of the simulation. Since this lost mass is not being re-accreted into the ISM, it is ejected from the halo. Although AGB elements show significantly different disk retention fractions, the CGM fractions are very similar to the CCSN elements. This is possibly because whatever AGB elements do end up in the halo were carried out through the same processes as the CCSN elements, leading to analogous evolution in the CGM.

\subsection{Stellar abundances}
\label{sec:stellar abundances}

\begin{figure}
  \centering
  \includegraphics[width=1.1\linewidth]{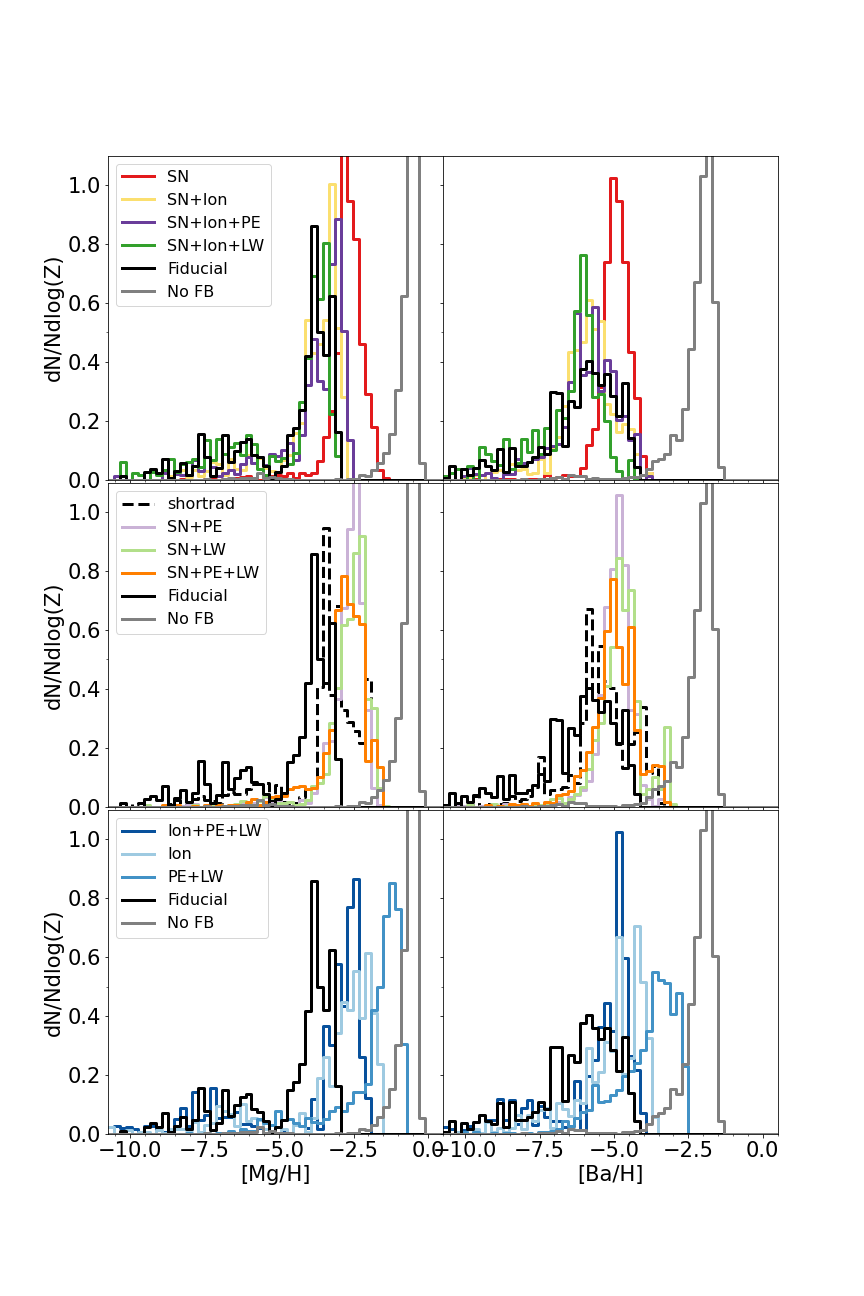}
  \caption{Stellar MDFs for all stars in a subset of our simulations at t = 500~Myr (for simulations with final run times less than 500~Myr, we only take those stars that would be alive at 500~Myr). For clarity, we do not show RPx0, RPx2, RPx5, or RPx10, which either show little variation with the fiducial run.}
  \label{fig:MDF1}
\end{figure}

The previous discussion focused on the time evolution of the instantaneous gas-phase abundances of the simulated galaxies, but the best observable of galactic chemical evolution in low mass dwarf galaxies is their stellar abundances. These are the convolution of the instantaneous gas-phase abundance distributions and the SFR. To examine the effect of feedback on stellar abundance patterns, we plot the normalized 1D metallicity distribution functions (MDFs) for the abundance ratios [Mg/H]\footnote{The notation [A/B] represents the abundance of element A relative to B, normalized to the solar ratio, in logscale: [A/B]~$\equiv$~log$_{10}$(N$_{\rm A}$/N$_{\rm B}$)~-~log$_{10}$(N$_{\rm A,\odot}$/N$_{\rm B,\odot}$).}, and [Ba/H] in Figure~\ref{fig:MDF1}.\footnote{ While our simulated galaxy is modelled after the $z=0$ properties of the Leo~P dwarf galaxy (see Paper~I), we emphasize that these stellar abundances are likely not representative of the actual abundance patterns in a Leo P like dwarf galaxy ($M_* \sim 10^6 M_{\odot}$) given that we only capture 1~Gyr of evolution, neglecting the possibility of pre-existing Type Ia SN progenitors. Thus,
we do not consider [Fe/H] or any similar tracer of Type Ia SNe enrichment, as the simulation has not been evolved for a long enough time for any element to be dominated by this channel (see Figure 2 of \citet{Emerick2018b}).} However, this provides insight into possible abundance variations in lower-mass dwarf galaxies that form a majority of their stars over timescales of $\sim$1~Gyr in the early Universe. These columns represent typical enrichment from CCSNe and s-process enrichment from AGB winds respectively. 

There are three general properties to compare across these MDFs: the location of the peak abundance, the width of the distributions, and the tails towards both higher and lower abundances. The increased peaks in nearly all simulations (compared to the fiducial), with the exception of those in the top panel with both SN and ionization feedback, are driven by the increase in ISM metals in each case, driven both by lower metal ejection fractions and higher SFR. These distributions are interesting with reference to Figure~\ref{fig:properties}. Simulations with similar $\langle Z_{\rm ISM}\rangle $ at the point of comparison here, for example, 500 Myr for the shortrad run and $\sim$350 Myr for SN+PE+LW (and also comparing to SN+PE and SN+LW, with slightly higher but still similar $\langle Z_{\rm ISM}\rangle $) can manifest larger offsets in the peak of their stellar MDFs. This suggests that there is a third mechanism with which stellar feedback regulates stellar abundances. The two obvious ones, as discussed previously, are: 1) by modifying the number of stars that form and thus the amount of metals produced and 2) by driving out some fraction of those metals from the ISM, but also, more subtly, 3) by determining how quickly ISM metals can be incorporated into new stars before being ejected. 

The radiation-only runs (bottom panels) also show the highest enrichment compared with the fiducial run, most significantly for PE+LW. Ion (only) and the Ion+PE+LW simulation exhibit more similarly peaked MDFs to the remaining SN runs.

\begin{figure}
  \centering
  \includegraphics[width=1.1\linewidth]{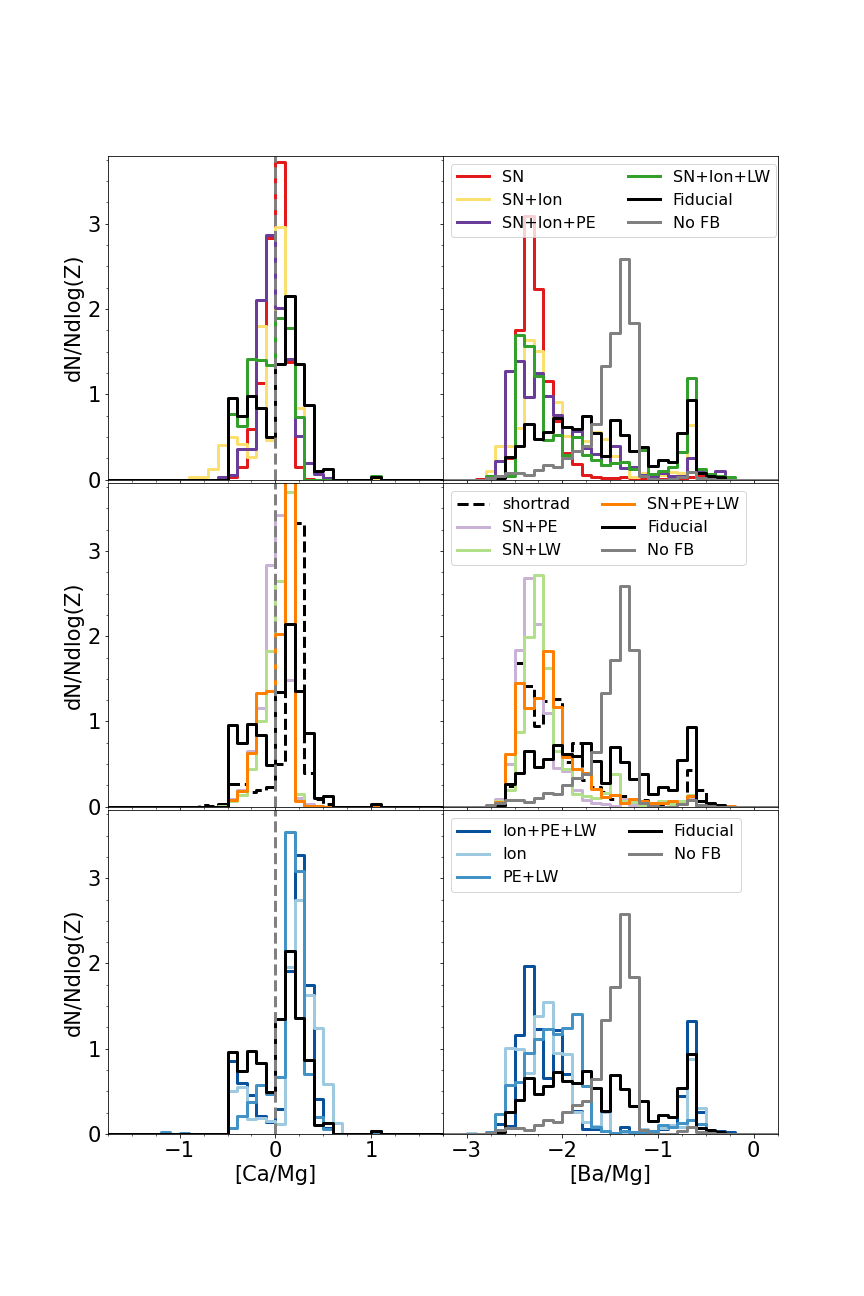}
  \caption{The same as Figure~\ref{fig:MDF1}, but for the abundance ratios [Ca/Mg], and [Ba/Mg]. Mg and Ca both trace enrichment from ccSNe, while and Ba traces enrichment from s-process via AGB winds. For clarity, we exclude NoFeed from [Ca/Mg] as it shows minimal variation with Ion, and exclude RPx0, RPx2, RPx5, and RPx10, which do not show significant variations with the fiducial run. The vertical grey line in the left panel denotes the solar [Ca/Mg] to emphasize the shift in typical [Ca/Mg] values -- which traces the mass dependence in our ccSNe yields-- with different sources of stellar feedback.}
  \label{fig:MDF2}
\end{figure}

While the distribution of stellar abundances in terms of total metal fractions ([X/H]) provides additional insight into the effects of feedback on galactic evolution, this does not necessarily provide additional constraints beyond the total metallicity alone. To examine if individual abundances could provide additional constraints, we plot two abundance ratios in Figure~\ref{fig:MDF2}, [Ca/Mg] and [Ba/Mg]. As shown, the largest differences across all runs lie in the widths of the distributions as compared to the fiducial simulation. In every case, the fiducial simulation produces a broader distribution for each abundance ratio. 

In [Ca/Mg], the additional star formation and enrichment in each scenario drives the MDF of the distribution closer to the IMF-averaged value of [Ca/Mg] for our yield set ([Ca/Mg] $\sim$ 0.22), homogenizing over some of the mass dependence between these two abundance ratios. This occurs most prominently for the radiation-only runs, and less so for the remaining runs. This is interesting, because Ca and Mg production varies with SN type in our yield model, with the most massive stars ($M_* = 25$~M$_{\odot}$) producing [Ca/Mg] = -2.96, and the least massive stars ($M_* = 8$~M$_{\odot}$) having [Ca/Mg] = 0.77 at stellar metal mass fractions $Z = 10^{-4}$.
The runs with peaks slightly less than the fiducial simulation have peak [Ca/Mg] closer to solar. This suggests a difference in how metals from the first vs.\ last SNe in a given star formation event are mixed in the ISM and released through outflows. This is an interesting signature of the effects of stellar feedback on stellar abundances beyond differences already shown between metals of different nucleosynthetic sources. However, determining this effect on mass-dependent SN yields conclusively requires further analysis, in particular the use of tracer particles or separate passive scalar fields in stellar mass bins to conclusively trace the evolution of elements from different SNe.

In [Ba/Mg], the fiducial simulation and those with both SNe and ionizing radiation exhibit broad, nearly flat MDFs across a large range ($> 2$ dex). The remaining runs all have more narrowly peaked (but still more broad than [Ca/Mg]) distributions around or below [Ba/Mg] of -2. The unifying origin of this difference is in how metals are retained or ejected across runs (Figure~\ref{fig:retention}). Those runs with narrower [Ba/Mg] distributions exhibit greater and more similar retention fractions of both species. It is precisely the relative difference in increased retention fractions between these two species that drives this broadening of these distributions (more retained Mg relative to Ba drives lower [Ba/Mg]). This indicates a second mode for increasing the width of abundance ratio spreads beyond the mixing differences discussed in Section~\ref{sec:metals} and previously in \citet{Emerick2018b}.

These results demonstrate that there are indeed differences in the resulting abundances for models with different stellar feedback, driven by how stellar feedback regulates the SFH of a galaxy relative to the enrichment timescales of different nucleosythnetic processes, and the retention / ejection of different elements in galactic winds.


\section{Discussion}

\subsection{Coupled Role of Supernovae and Ionizing Radiation}

It is tempting to try to answer the question ``What is the dominant source of stellar feedback in galaxy evolution?``, yet it has become increasingly apparent \citep[e.g.][]{Agertz2013,Agertz2020,Hu2016,Hopkins2018,Smith2019} that this is not a well-posed question. In the context of a low-mass dwarf galaxy, SN feedback may be considered to be the most important single source of feedback in terms of its ability to both regulate star formation, drive outflows, and generate turbulence needed for metal mixing in the ISM, this work shows clearly that it is only in combination with additional sources of feedback that SNe have such a significant effect. Of the processes examined here -- SNe, photoelectric heating, LW radiation, ionizing radiation, and radation pressure -- it is the combined effects of SNe and stellar ionizing radiation that could be considered to be the dominant sources of stellar feedback. 

\subsection{Non-Ionizing Radiation}

While PE heating of dust grains is an important source of heating in dense gas of more metal-rich environments like the Milky Way ISM \citep[e.g.][]{BakesTielens1994,Wolfire2003}, its role in low-metallicity dwarf galaxies remains less clear. Recently, \cite{Hu2017} found PE heating to be unimportant in a galaxy with stellar mass $\sim$10$^{7}$~M$_{\odot}$ and $Z\sim$0.1~Z$_{\odot}$, in contrast to \cite{Forbes2016} who found it to be a significant source of feedback in their dwarf galaxy of similar mass and metallicity.\footnote{\cite{Hu2017} argue the difference was due to a different treatment of metal line cooling in self-shielded gas.} Our results here clearly support the conclusion of \citet{Hu2017} that PE heating is not a significant source of stellar feedback in the low-metallicity regime of our low-mass dwarf galaxy. There is no significant difference between the SN and SN+PE and the SN+Ion and SN+Ion+PE runs for the properties studied here. 

Far more important, however, is LW radiation, which does have a significant effect through regulating the H$_2$ content of this galaxy. As H$_2$ is a dominant coolant in this low-metallicity regime, this has the effect of preventing some gas from cooling and collapsing to form stars, but these effects are sub-dominant to effects of stellar ionizing radiation as discussed above.

\subsection{Radiation Pressure}
\label{sec:radiation pressure}

The importance of radiation pressure to galaxy evolution in various regimes is also debated \citep[e.g.][]{HQM2011,HQM2012,Agertz2013,Ceverino2014,Sales2014,Rosdahl2015,Kannan2020}, with detailed analysis and discussions presented in \citet{Krumholz2018} and \citet{Hopkins2020a}). In the low-mass dwarf galaxy regime, radiation pressure is sub-dominant to photoheating and ionization, but can still play a significant role in star formation regulation \citep{WiseAbel2012}. We focus here on the role of radiation pressure in the single-scattering limit, ignoring the potential increase in deposited momentum from resonant re-scattering of UV photons into the IR as that appears to be a limited enhancement \citep{Reissl2018}. Overall, we find little variation in the properties of our low-mass dwarf galaxy when varying the radiation pressure factor from 0 (off) to 10. However, we do find a slight trend of increasing mass loading factor near the galaxy (0.1~$R_{\rm vir}$) with increasing radiation pressure factor due to the increased pressure on neutral gas near the disk. However, the radiation pressure factor used here is an ad hoc way to explore how additional momentum transfer from multiple-scattering off of dust in the IR may affect the evolution. This trend with $\eta_M$ would be unlikely to occur in a more self-consistent simulation that directly tracks multiple-scattering, as there is likely little-to-no dust external to this already metal-poor galaxy.  


As this is only a single, isolated, low-mass dwarf galaxy we cannot universally say that radiation pressure is subdominant in this regime, but it appears to be unlikely for galaxies much like ours. This is in disagreement with the results from \citet{WiseAbel2012}, who used the same radiation feedback methods. However, that work studied star formation across a population of dwarf galaxies with different stellar masses and morphologies than the galaxy examined here. This conclusion may also be resolution dependent; higher resolution simulations capable of resolving higher gas densities may see a larger impact from adjusting the effectiveness of radiation pressure \citep[see][]{Krumholz2018}.


%
%
%
%
%
%

\section{Conclusion}

In a set of simulations studying the evolution of a low-mass, isolated dwarf galaxy modelled after the $z = 0$ properties of the Leo~P dwarf galaxy, we explore the effects that each channel of a star-by-star model for multi-channel stellar feedback has on the galaxy's evolution. We test the role that PE heating, LW radiation, ionizing radiation, radiation pressure, and SNe have on this galaxy through seventeen different simulations at moderately high (3.6~pc) resolution. We find that each form of feedback can generate some form of self-regulated star formation in our dwarf galaxy. However, the different combinations of these processes that we tested produce qualitatively different star formation rates and galaxy properties. Even in cases where the total stellar mass formed is similar across runs, the different feedback channels can produce significant changes in other galaxy properties, particularly in the metal content of the ISM. We pay particular attention to their ability to drive metal-enriched outflows, metal mixing, and stellar abundance patterns of different elements in our galaxy.

\subsection{Importance of Different Stellar Feedback Processes}

Due to the complex interplay of different forms of stellar feedback, it is an ill-posed problem to inquire which one is the most important or effective. However, we can draw conclusions from the relative effects on galaxy evolution in models that do or do not include certain feedback processes.

We find that SNe and ionizing radiation \textit{together} are the dominant sources of stellar feedback regulating star formation, the multi-phase ISM, and the driving of outflows in our model of a low-mass dwarf galaxy. These feedback channels act in combination, with ionizing radiation decreasing the ambient ISM density in which SNe explode and carving the channels of warm, diffuse gas needed to vent significant amounts of mass, metals, and energy
in outflows. All of our feedback models generate outflows dominated by cool mass and metal loading, in contrast to the hotter outflows in more massive galaxies; this is less true for energy loading, with more comparable hot and cold energy loading factors in some runs, and larger hot loading factors in runs with $\Sigma_{\rm SFR} > 10^{-4}$~M$_{\odot}$~kpc$^{-2}$~yr$^{-1}$. In general, including ionizing radiation feedback produces cooler mass and metal outflows while increasing the total mass and metal loading factors (more efficient feeedback). 

We find that the destruction of H$_2$ through LW radiation is an important source of feedback in our low-metallicity dwarf galaxy. In this regime, H$_2$ is the dominant coolant. LW radiation produces a far greater effect on the star formation and outflow properties of our dwarf galaxy than PE heating from FUV radiation, in part also due to the limited dust content in this low-metallicity environment.

We additionally explore the effects of radiation pressure and find that an order of magnitude boost in the radiation pressure effectiveness in the single-scattering limit is necessary to noticeably affect the evolution of the galaxy, and even then the trend as a function of this boost factor is not clear. We conclude that radiation pressure is not a signficant source of feedback for this particular galaxy.

\subsection{Feedback and Galactic Chemical Evolution}

In driving metal-enriched outflows, the presence of stellar feedback changes how metals are retained by and mixed within our galaxy. This manifests itself across simulations as a difference in the ISM metal abundances and abundance spreads, which can have a significant impact on the stellar abundance patterns in this galaxy. This is predominantly observable by examining the total abundance of a given element, but also imprints itself upon the abundance ratios in stellar abundance space, albeit to a smaller degree. As we found in \citet{Emerick2018b}, elements released in SNe vs. AGB winds are ejected at larger rates and mix more efficiently in our fiducial simulation. The difference between the behavior of these two elements decreases in runs with fewer forms of feedback than our fiducial run, with turning off ionization and SNe having the largest effects (in order). We find that runs without ionizing radiation (and also without SNe) produced significantly narrower abundance spreads in [Ba/Mg] (proxies of AGB and core collapse SNe elements, respectively) than our fiducial simulation. In addition, each simulation shows differences in the final stellar [Ca/Mg] MDFs---both released in SNe, but with a ratio that is strongly dependent on stellar mass. The particular choice of stellar feedback model---when coupled to a model for mass and metallicity dependent stellar yields---can produce qualitatively different stellar abundance ratios. 

In summary, this analysis shows that including a varying set of stellar feedback modes can drive significantly different star formation, outflow, and chemical abundance properties in low mass dwarf galaxies. While SNe and ionizing radiation have the most dramatic effects in this regime, particularly in combination, LW radiation is also important. Feedback's ability to drive metal mixing in the ISM and metal enriched outflows, and how it does this differently for elements from distinct sources (e.g. AGB winds vs. SNe) influences the resulting stellar abundance patterns of our dwarf galaxy. This points to the importance of comparing simulations of galaxy evolution to the detailed stellar abundance patterns in observed galaxies as a particularly strong constraint on the underlying physics. This would be best done in high resolution cosmological simulations of galaxy evolution, capable of capturing the details of halo growth and accretion in more realistic environments than the idealized isolated galaxy presented here.

\section*{Acknowledgments} GLB acknowledges support from NSF grants AST-1615955 and OAC-1835509 and NASA grant NNX15AB20G. M-MML was partly supported by NSF grant AST18-15461. We gratefully recognize computational resources provided by NSF XSEDE through grant number TGMCA99S024, the NASA High-End Computing Program through the NASA Advanced Supercomputing Division at Ames Research Center, Columbia University, and the Flatiron Institute. This work made significant use of many open source software packages. These are products of collaborative effort by many independent developers from numerous institutions around the world. Their commitment to open science has helped make this work possible. 

\software{\textsc{yt} \citep{yt}, \textsc{Enzo} \citep{Enzo2014}, \textsc{Grackle} \citep{GrackleMethod}, \textsc{Python} \citep{VanRossum1995python}, \textsc{IPython} \citep{perez2007ipython}, \textsc{NumPy} \citep{oliphant2006guide}, \textsc{SciPy} \citep{SciPy}, \textsc{Matplotlib} \citep{hunter2007matplotlib}, \textsc{HDF5} \citep{Fortner1998HDF,Koranne2011}, \textsc{h5py} \citep{h5py}, \textsc{Astropy} \citep{astropy:2013,astropy:2018}, \textsc{Cloudy} \citep{Cloudy2013}, and \textsc{deepdish}}




\appendix 
\setcounter{section}{0}
\counterwithin{figure}{section}

\section{Morphological Properties}

\label{appendix:panels}
In Section~\ref{sec:morphology}, we discuss the morphology of our galaxy early in its evolution (125~Myr), just after the initial star formation burst for each simulation and at a time reached by each simulation. We explore here the gas morphology at later, more quiescent times, which are not reached by all simulations.

\begin{figure*}
  \centering
  \includegraphics[width=0.95\linewidth]{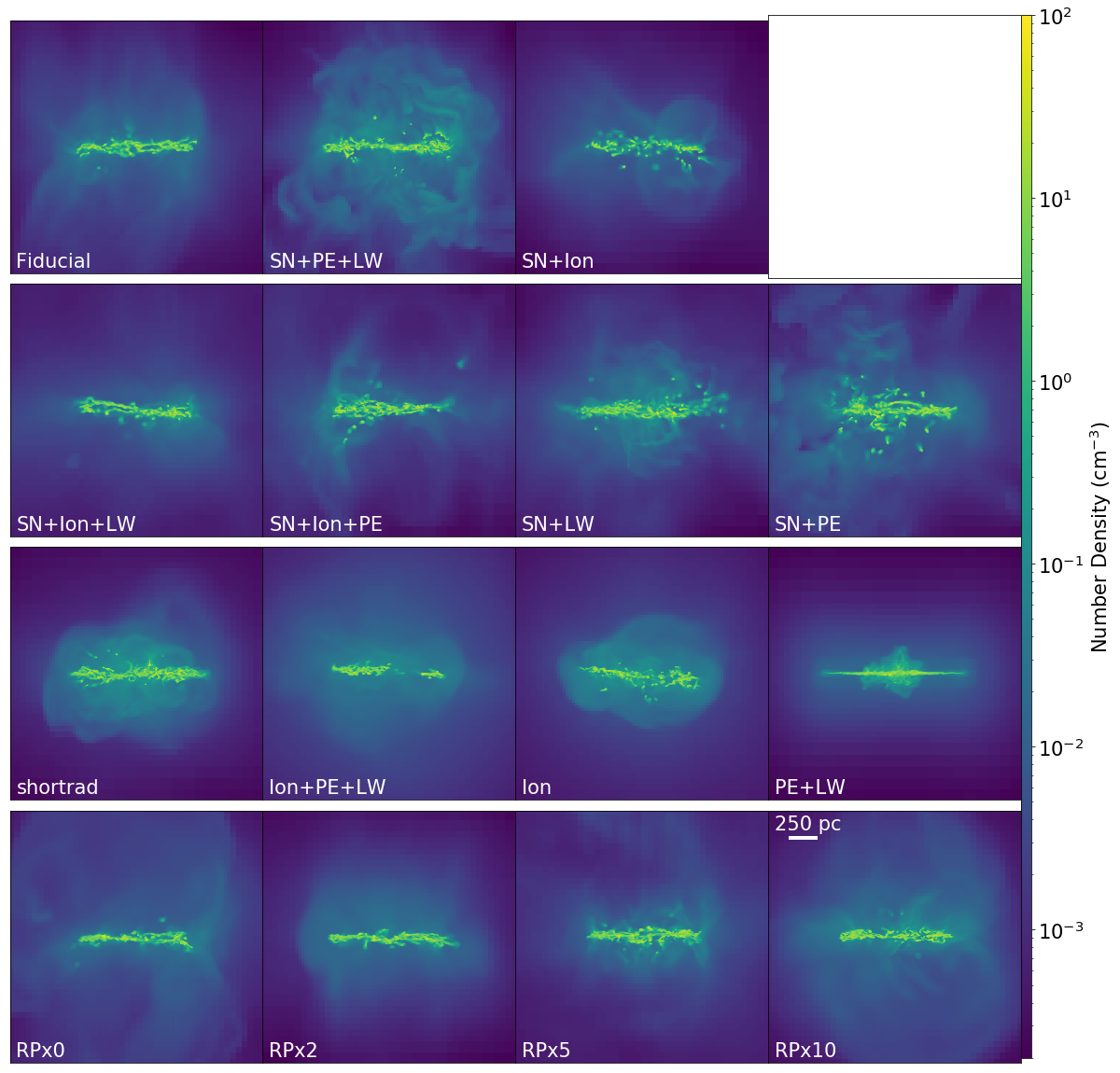}
  \caption{Edge-on projections of gas number density for each of our runs with feedback at time $t=350$~Myr for all but the SN-only simulation.}
  \label{fig:panel_plot_3}
\end{figure*}

\begin{figure*}
  \centering
  \includegraphics[width=0.95\linewidth]{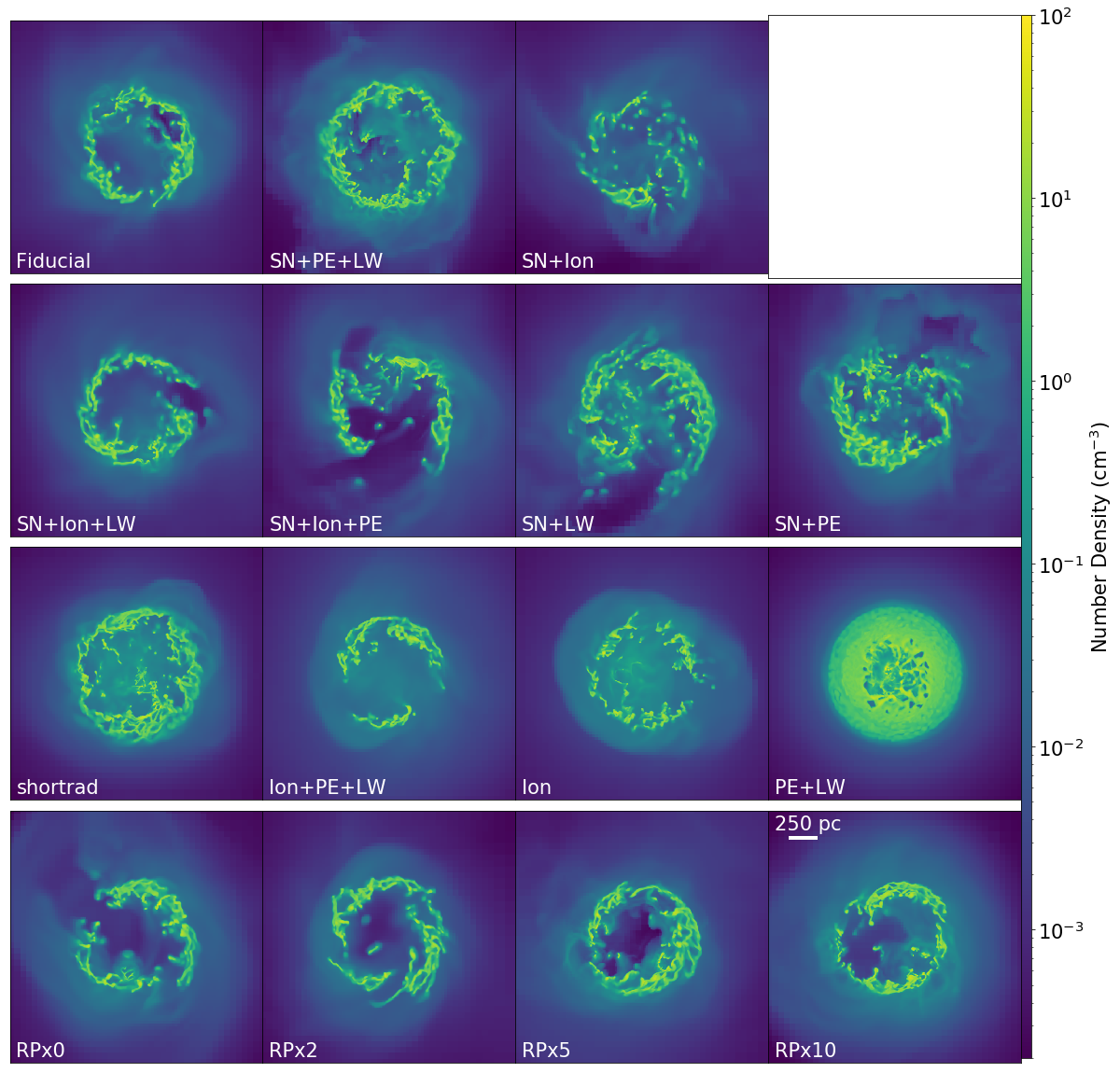}
  \caption{Same as Figure~\ref{fig:panel_plot_3}, but face-on.}
  \label{fig:panel_plot_4}
\end{figure*}

In Figure~\ref{fig:panel_plot_3} and Figure~\ref{fig:panel_plot_4} we show edge-on and face-on projections of gas number density at 350~Myr for each simulation. In the edge-on panels, the same general trends seen in Section~\ref{sec:morphology} across runs can be seen. The largest difference in each panel is the decrease in each case of the amount and morphology of gas beyond the disk of the galaxy. During this period of low SFR for all but the shortrad simulation, the extraplanar gas around the disk is more uniform and less extended than during times of higher SFR. 

This is visible as well in the face-on panels, which still exhibit similar lower-density cores and higher-density rings to those at t=125~Myr. The key difference is that the lower-density, inner regions have a higher typical density than when the galaxy was actively star forming. In addition, at 350~Myr, the outer rings in each galaxy tend to be clumpier and less continuous, broken up by the star formation and feedback that occurred over the preceding 250~Myr. PE+LW shows the least variability between these two times, and also has the most uniform SFR of all the simulations.

%
%

\bibliographystyle{yahapj}
\bibliography{refs}

\end{document}